\documentclass{aa}  

\usepackage{graphicx}
\usepackage{txfonts}
\usepackage[pdfpagelabels=false]{hyperref}	% Hyperlinks
\hypersetup{colorlinks=true,linkcolor=blue,citecolor=blue,filecolor=blue,urlcolor=blue,}
\makeatletter
\renewcommand*\aa@pageof{, page \thepage{} of \pageref*{LastPage}}
\makeatother

\usepackage{amsmath}	% Advanced maths commands
\usepackage{amssymb}	% Extra maths symbols
\usepackage[compatibility=false]{caption}
\usepackage{subcaption}
\usepackage{xcolor}
\usepackage{placeins}
\newcommand\molH{\text{H\textsubscript{2}}}
\newcommand\water{\text{H\textsubscript{2}O}}
\newcommand\term[3]{\text{\textsuperscript{#1}#2\textsubscript{#3}}}
\newcommand\dd{{\rm d}}

\definecolor{orange}{rgb}{1,0.5,0}
\definecolor{sred}{rgb}{.5,0,0}

\begin{document}

   \title{Photoevaporation of protoplanetary discs with PLUTO+PRIZMO}

   \subtitle{I. Lower X-ray--driven mass-loss rates due to enhanced cooling}
   \titlerunning{X-ray photoevaporation with PLUTO+PRIZMO}

   \author{A. D. Sellek\inst{1,2}
          \and
          T. Grassi\inst{3}
          \and
          G. Picogna\inst{4}
          \and
          Ch. Rab\inst{3,4}
          \and
          C. J. Clarke\inst{2}
          \and
          B. Ercolano\inst{4,5}
          }

   \institute{Leiden Observatory, Leiden University, 2300 RA Leiden, The Netherlands\\
              \email{sellek@strw.leidenuniv.nl}
             \and
              Institute of Astronomy, University of Cambridge, Madingley Road, Cambridge CB3 0HA, UK
             \and
              Max-Planck-Institut für extraterrestrische Physik, Giessenbachstrasse 1, D-85748 Garching, Germany
             \and
              University Observatory, Faculty of Physics, Ludwig-Maximilians-Universität München, Scheinerstr. 1, D-81679 Munich, Germany
             \and
              Exzellenzcluster “Origins,” Boltzmannstr. 2, D-85748 Garching, Germany
             }

   \date{Received March 28, 2024; accepted July 31, 2024}

% \abstract{}{}{}{}{} 
% 5 {} token are mandatory
 
  \abstract
  % context heading (optional)
  % {} leave it empty if necessary  
   {Photoevaporation is an important process for protoplanetary disc dispersal but there has so far been a lack of consensus from simulations over the mass-loss rates and the most important part of the high-energy spectrum for driving the wind.}
  % aims heading (mandatory)
   {We aim to isolate the origins of these discrepancies through carefully-benchmarked hydrodynamic simulations of X-ray photoevaporation with time-dependent thermochemistry calculated on the fly.}
  % methods heading (mandatory)
   {We conduct hydrodynamic simulations with \textsc{pluto} where the thermochemistry is calculated using \textsc{prizmo}. We explore the contribution of certain key microphysical processes and the impact of using different spectra used previously in literature studies.}
  % results heading (mandatory)
   {We find that additional cooling results from the excitation of O by neutral H, which leads to dramatically reduced mass-loss across the disc compared to previous X-ray photoevaporation models, with an integrated rate $\sim 10^{-9}\,{\rm M}_{\sun}\,\mathrm{yr^{-1}}$. Such rates would allow for longer-lived discs than previously expected from population synthesis. An alternative spectrum with less soft X-ray produces mass-loss rates around a factor of 2-3 times lower. The chemistry is significantly out of equilibrium, with the survival of H\textsubscript{2} into the wind aided by advection. This leads to its role as the dominant coolant at 10s au - thus stabilising a larger radial temperature gradient across the wind - as well as providing a possible wind tracer.}
  % conclusions heading (optional), leave it empty if necessary 
   {}


   \maketitle
%
%-------------------------------------------------------------------
\section{Introduction}
Protoplanetary discs are the site of planet formation, as evidenced by direct observations of protoplanets around the T Tauri star PDS70 \citep{Keppler_2018,Muller_2018} as well as candidates around the more massive Herbig stars AB Aurigae b \citep{Currie_2022} and HD 169142 \citep{Gratton_2019,Hammond_2023}. The formation, growth, and migration of planets in protoplanetary discs is ultimately limited by disc dispersal, thought to happen rapidly \citep{Skrutskie_1990,Simon_1995,Wolk_1996} after around 2-8 Myr of evolution \citep{Pfalzner_2022} depending on the star-forming environment \citep{Michel_2021} and stellar mass \citep{Ribas_2015}.

Disc winds are widely evidenced throughout the disc lifetime \citep{Pascucci_2023} and are one important factor contributing to the evolution and dispersal of these discs. Typically two paradigms are considered: photoevaporation, where thermal pressure gradients resulting from heating by high-energy stellar radiation drive the wind, and magnetically-driven winds, where gradients in the magnetic pressure and centrifugal forces (resulting from material being forced to co-rotate with the magnetic field) launch gas outwards \citep[although][define the intermediate case of magnetothermal winds as those where both magnetic and thermal effects play a role]{Bai_2017}.

Photoevaporation is a particularly attractive paradigm for the late-stage dispersal of discs.
The mass-loss rates that can be produced by cold magnetically-driven wind models powered by accretion become less significant as the disc accretion rate drops. Therefore, these winds become more tenuous and hence more optically thin to high-energy radiation, allowing it to more easily reach the disc and drive photoevaporation \citep{Kunitomo_2020,Weder_2023}.
On the other hand, photoevaporation rates are largely insensitive to the properties of the underlying disc (at least until any transparent gas cavity opens): the disc adjusts to deliver the mass flux expected from the thermochemistry of the wind \citep{Owen_2012} and the mass-loss rate remains high until the disc becomes optically thin to the dominant radiation field driving photoevaporation.

Photoevaporation also explains a number of observational constraints which suggest dispersal progresses from the inside out. The ``UV switch'' model of \citet{Clarke_2001} shows how photoevaporation should carve a gap, and then a cavity, inside the inner few au.
Such a mode of dispersal is supported by the relative disc lifetimes at different wavelengths \citep{Ribas_2014} and the evolutionary loci of discs in colour-colour space \citep{Koepferl_2013}, which imply the loss of the warmer inner material - which emits at shorter wavelengths - first.

However, theoretical predictions for the strength of photoevaporative winds lack consensus. Mass-loss rates in the literature have varied from $10^{-10}-10^{-9}\,{\rm M}_{\sun}\,\mathrm{yr^{-1}}$ for models driven by Extreme Ultraviolet (EUV) \citep{Hollenbach_1994} to $10^{-8}-10^{-7}\,{\rm M}_{\sun}\,\mathrm{yr^{-1}}$ for X-ray--driven models \citep{Owen_2010,Owen_2011,Owen_2012,Picogna_2019,Ercolano_2021,Picogna_2021} or models driven by Far Ultraviolet (FUV) \citep{Gorti_2015,Nakatani_2018a,Nakatani_2018b}.

In part, these differences may be attributed to different methodologies - since the problem of thermochemistry on the fly with hydrodynamics is very expensive, different works have compromised on different factors.
For example, \citet{Hollenbach_1994} or \citet{Gorti_2009a,Gorti_2015} do not explicitly solve the hydrodynamics, but estimate what solution should result given their calculated temperature structure. On the other hand, X-ray photoevaporation models since \citet{Owen_2010} have typically solved the hydrodynamics with temperatures prescribed as a function of the ionisation parameter $\xi=\frac{L_X}{nr^2}$ \citep{Tarter_1969}, which roughly speaking represents the ratio of the X-ray photoionisation heating to the two-body collisional cooling. Such an approach allows a large parameter space to be explored at the expense of not being able to consider effects such as out-of-equilibrium thermochemistry or adiabatic cooling resulting from the PdV work done by the expanding gas. The $\xi-T$ relationships have also been calculated assuming purely atomic gas using the \textsc{mocassin} code \citep{Ercolano_2003,Ercolano_2005,Ercolano_2008} and only EUV/X-ray radiation.
In recent years, there have been moves towards on-the-fly thermochemistry and hydrodynamics both for photoevaporation \citep[e.g.][]{Wang_2017,Nakatani_2018a,Nakatani_2018b,Komaki_2021} and also for magnetohydodynamic (MHD) winds \citep{Wang_2019,Gressel_2020}, but such efforts naturally come at the expense of solving the thermochemistry in detail and hence are very sensitive to the simplifying assumptions made.

The fact that, typically, several things - such as the irradiating spectrum, possible coolants, radiative transfer method, grid extent, explicit calculation of hydrodynamics, and assumption of thermochemical equilibrium - differ between each methodology, has made it very difficult to make direct comparisons between the above works and clearly assess the dominant factors responsible for the divergent results.
%As a result of these different methodologies, many works lack critical things that may be of importance. 
For example, \citet{Wang_2017} - who favoured a low-mass-loss-rate EUV-driven wind - attributed the higher mass-loss rates of X-ray models to the lack of molecular cooling. On the other hand, \citet{Sellek_2022} demonstrated that a lack of effective optical line cooling in the work by \citet{Wang_2017} meant that their predicted wind temperatures were too high so despite the additional molecular cooling, this thermochemical treatment was still not sufficiently complete. More importantly, \citet{Sellek_2022} showed that the results were very sensitive to the shape of the X-ray spectrum, finding that the simplified spectrum used by \citet{Wang_2017} where they assumed all the X-rays to have an energy of $1\,\mathrm{keV}$ meant that the X-rays interacted too weakly with the gas to achieve significant heating; that is to say \citet{Wang_2017} were prevented from finding an X-ray wind primarily by their lack of soft X-rays.

In this work, we seek to build the most comprehensive photoevaporation model to date, using \textsc{prizmo} \citep{Grassi_2020} to combine hydrodynamics with thermochemistry calculated on the fly; for this first application, we focus on X-ray--driven winds.
We stress the importance of making sure that our approach is sufficiently complete to capture thermochemistry in all possible wind physical environments, including ionised, neutral atomic, and molecular gas. As such, we emphasise benchmarking the temperatures calculated by \textsc{prizmo} against previous models and other codes. Important aspects of this include explicitly exploring the role of what we find to be the most critical thermochemical processes, and enabling direct comparison to other models at each stage with appropriate spectra.

Section \ref{sec:methods_thermo} sets out the key features of \textsc{prizmo} and any choices we make for the thermochemistry, while Sect. \ref{sec:methods_hydro} explains how we use \textsc{prizmo} in conjunction with \textsc{pluto} for the hydrodynamics. Section \ref{sec:benchmarks} presents benchmarks of the temperature structure of winds in different regimes to illustrate the key successes of \textsc{prizmo} and elucidate important differences to previous works.
We then demonstrate that hydrodynamic simulations using \textsc{prizmo} substantially alter the temperature profile and reduce the mass-loss rates of X-ray-driven winds in Sect. \ref{sec:results}. We also derive the corresponding mass-loss profile, establish the importance of non-equilibrium thermochemistry, and explore how the mass-loss rate is affected by the microphysics and irradiating spectrum. In Sect. \ref{sec:discussion}, we briefly discuss the implications of these results for simulations of photoevaporation, disc evolution and observations of disc winds. We summarise our conclusions in Sect. \ref{sec:conclusions}.

%--------------------------------------------------------------------
\section{Methods: thermochemistry}
\label{sec:methods_thermo}
Previous works have highlighted the need for an approach to photoevaporation modelling which a) uses a broad, detailed, spectrum for the radiation, b) includes sufficient atomic transitions to provide cooling regardless of the temperature (and ionisation), c) includes cooling of molecules with abundances calculated self-consistently with the local radiation field, d) is done on the fly with hydrodynamics to avoid assuming radiative thermal equilibrium in the regions where adiabatic cooling may be prevalent.

Such a comprehensive approach demands a fast code that can solve the thermochemistry - a term here used to describe collectively the heating and cooling that results from the chemical composition of the gas and interactions between its constituents - across molecular and atomic regimes and can be called as a library by hydrodynamics codes using the operator splitting technique \citep[e.g.][]{Teyssier_2015}.
\textsc{prizmo} \citep{Grassi_2020} was developed specifically for this purpose.
\textsc{prizmo} has been significantly updated since its first release, and the main changes are described in Sect. \ref{sec:prizmo_updates}\footnote{This version of \textsc{prizmo} is available from GitHub at: \url{https://github.com/tgrassi/prizmo}, commit \texttt{b2158ae}.}.

%First, we provide an overview of the code and its preprocessor inputs. 
%Ultimately, this must include both gas-phase chemistry (typically two-body reactions) - and \textit{photochemistry} - reactions such as photoionisation and photodissociation of a single atom/ion/molecule induced by the local radiation field - in order to know the gas composition.

\textsc{prizmo} is configured by means of a \textsc{python} preprocessor to prepare the appropriate thermochemistry \textsc{fortran} modules and corresponding required data files.
These modules include \textsc{C} wrappers, allowing them to be compiled with codes such as \textsc{pluto} (Sect. \ref{sec:methods_hydro}). 

The key inputs to the preprocessor are:
\begin{itemize}
    \item Atomic data (level energies, degeneracies, Einstein coefficients, and fits to collisional de-excitation rates)
    \item A list of (photo)chemical reactions with rate coefficient expressions
    \item A user-defined spectrum and number of energy bins 
    \item Frequency-dependent dust optical constants
\end{itemize}
Our preprocessor inputs are detailed in Sections \ref{sec:network}-\ref{sec:dust}.

A key feature of \textsc{prizmo} is its multi-frequency approach: all photochemical cross-sections and dust opacities (which are integrated over a grain size distribution) resulting from the selected reaction network and dust optical constants are tabulated as a function of photon energy. All the radiation-related processes (e.g., photoheating) therefore use the same multi-frequency approach where possible. This significantly improves the consistency of the results, by allowing \textsc{prizmo} to calculate photochemistry using the local radiation field (rather than assuming rates based on a standard field).

\subsection{Updates to \textsc{prizmo}}
\label{sec:prizmo_updates}

\subsubsection{Structure/architecture}
Since \citet{Grassi_2020}, \textsc{prizmo} has been modified to reduce the global computational time and improve the code readability, and further modifications were made to the physics and chemistry involved in the present application.
Nevertheless, despite these improvements affording an approximately 50\% speed-up, the thermochemistry still dominates the execution time and is expected to typically be around 100 times longer than would be required to have no additional impact compared to pure hydrodynamics. This section describes the improvements from \citet{Grassi_2020}. 
%The coding improvements allow approximately a 50\% speed-up.%, obtaining around a microsecond average integration time per cell.
%Nevertheless, this is still far from the target of tens of nanoseconds required

\subsubsection{Rate coefficients calculation}
We increased the vectorisation of rate coefficient calculations in the chemical network by grouping chemical rate coefficients by the same arithmetic expression, for example $\alpha (T/300\,{\rm K})^\beta\exp{(-\gamma/T)}$ and $\alpha(T/300\,{\rm K})^{\beta}$ belong to different groups. Since most of the rates have relatively simple algebraic forms, with this approach we obtain a speed-up similar to interpolating the rate coefficients in temperature in the log-log space.

We have also approximated some rates in order to precompute common algebraic quantities; for example, $(T/300\,{\rm K})^{-0.49}$ has been approximated to $(T/300\,{\rm K})^{-0.5}$, which is then a common term to several reaction rates, and hence can be precomputed. 
These small rate variations do not impact the final results, and have a $\sim1-2\%$ reduction on the computational time. However, it is important to notice that small speedups have a relatively large impact on simulations that require thousands of core hours. 
\subsubsection{Atomic cooling solver}
\label{sec:methods_atomicSolver}
The linear solver in the atomic cooling routine has been extended to solve linear systems analytically up to 5$\times$5 matrices without using \textsc{Lapack}. This considerably improved the speed of the atomic cooling linear solver for the larger atomic line systems needed to accurately capture atomic cooling across the range of temperatures found in protoplanetary discs and winds \citep[see Sect. \ref{sec:methods_atomic}, also][]{Sellek_2022}.
However, the atomic cooling routine occupies a large fraction of the computational time, mainly due to fitting functions in the log-log space employed to evaluate the rate coefficient with collisional partners. Fits have a relatively large computational impact because they require $10^{\rm x}$-type operations, where $x$ is a double precision number. We reduce the impact of their original algebraic functions by vectorising fitting operations over temperature. In addition to this, the combination of rate interpolations and solving the $N\times N$ linear system analytically provides a good computational speed with respect to, for example, multidimensional linear fits that require a dimension for each collisional partner and for temperature and, analogously to the other method, employ $10^{\rm x}$-type operations.

\subsubsection{Photoionisation}
Photoionisation is calculated using the cross-sections of \citet{Verner_1996} for each of the neutral atomic species in our network (H, He, C, O).
Modelling each species separately - rather than using a single approximated total photoionisation cross-section - is necessary to follow the ionisation balance of each species individually and self-consistently with the heating. It also allows flexiblity in treating variations in the gas compositions, for example due to altered elemental abundances or regions of enhanced ionisation.
However, note that as well as due to the direct radiation field of the star, ionisation can occur as a result of the diffuse radiation field created by recombinations. Since we only conduct radial ray tracing in this work, we cannot follow these photons accurately and hence assume they are reabsorbed on the spot. Although the diffuse field of one element can potentially ionise those with a lower ionisation potential (e.g., ground-state recombinations of He may ionise H), for simplicity we assume that the reabsorption occurs by the same element, noting that this may somewhat underestimate the energy liberated into the gas by photoelectrons but this will only make any significant difference in regions dominated by EUV heating while it is a much smaller correction if the original ionisation was by an X-ray photon.

Moreover, since doubly-ionised and higher species were not included in the network we use in this work (see Sect. \ref{sec:network}), we cannot include here photoionisation reactions of the singly ionised species. Similarly, neither can the code account for multiple ionisation by the Auger-Meitner effect, although the cross-section due to ionisation of inner (K) shell electrons is included \citep{Verner_1995} and is assumed to always result in a single electron being ejected.
Consequently, the entire yield over the (outer shell) ionisation energy was available to contribute to heating (assuming that any excited products from inner shell ionisation de-excite rapidly through collisions).
We do not produce secondary electrons, either by the Auger effect or by rapid collisional ionisation, save for what may already be produced by the thermalised electrons in the network. This may overestimate the heating by a small factor \citep[e.g.]{Maloney_1996}.
The efficiency of the heating is thus 100\% for photoionisation of completely neutral gas and 0\% for photoionisation of fully (singly-)ionised gas.
We intend to address these limitations in future works.

Neither X-ray photoionisation of molecules nor of dust is included in the present code (while the cross-section of molecules can be well-approximated by the sum of that of the atoms, the branching ratios of the outcomes are poorly known).

\subsubsection{Molecular cooling and heating}
It has been suggested that a key process missing in previous X-ray photoevaporation calculations that relied on \textsc{mocassin} was molecular cooling \citep{Wang_2017}.
Hence, we include it here according to tables computed from \citet{Omukai_2010} for CO cooling and from the piece-wise functions of \citet{Glover_2008,Glover_2015} for \molH{} cooling. While we do not include cooling from \water, we do not see it at sufficient levels in the wind for it to be an important coolant.

Molecules can also contribute to heating: photodissociation and pumping of \molH{} are included, each proportionally to the photodissociation rate calculated by the chemical network.

\subsubsection{Photoelectric heating}
We have changed the photoelectric routine following \citet{Kamp_2000} as reported in \citet{Woitke_2009}, their Eq.\,(93), that is related to the strength of the FUV radiation field $\chi$. This relation is less accurate and tunable than the full formalism of \citet{Weingartner_2001} and \citet{Weingartner_2006}, but we found it to be less prone to show unphysical heating when coupled with the radiative transfer, mainly due to implementation details.
Note that this formulation gives only the heating from silicate grains; in the current code no heating from Polyaromatic Hydrocarbons (PAHs) is included. PAH heating can be of consequence in the upper layers of the disc at 10s au and in the outer disc \citep{Woitke_2009} and while this shouldn't affect the models presented in this paper (which include little FUV in the spectra), it will need improving in future work where we aim to apply the code to FUV photoevaporation.   

\subsubsection{Dust temperature and dust cooling}
The previous code version computed the dust temperature using a bisection method applied to Eq.\,(41) in \citet{Grassi_2020}. We obtain faster integration times by linearly interpolating a pre-generated table of the form $T_{\rm d}=f(\Gamma_{\rm a}, T, n)$, where $\Gamma_{\rm a}$ is the absorption integral using the impinging radiation, as described in Eq.\,(43) of 
\citet{Grassi_2020}, $T$ is the gas temperature, and $n$ is the total gas density. Analogously, we compute the dust cooling and heating with the same approach using two pre-computed separated tables. Using two tables allows dividing heating (positive) and cooling (negative) in the log-log space. 

\subsection{Species \& reactions in our network}
\label{sec:network}
Our network is based around the elements H, He, C, and O with interstellar medium (ISM) gas-phase abundances consistent with previous photoevaporation models \citep[e.g.][]{Ercolano_2009,Wang_2017}: $\mathrm{He/H}=0.1$, $\mathrm{C/H}=1.4\times10^{-4}$ and $\mathrm{O/H}=3.2\times10^{-4}$ \citep{Savage_1996}.

We use the chemical network described in \citet[][Appendix\,G]{Grassi_2020}, which is designed to obtain the correct temperature structure, rather than the correct chemical structure of the less abundant chemical species.
It is loosely based on Photodissociation Region (PDR) chemistry and includes the neutral and singly-ionised versions of each atom, electrons, 22 neutral/ionised gaseous molecules, and finally CO and \water{} ices, all for a total of 33 species.
%While not all significant in their own right, several molecules are important intermediaries, for example CH\textsubscript{2}\textsuperscript{+} and CH in CO production.
A total of 290 reactions are included, covering two-body gas-phase reactions, photoionisation, photodissociation, cosmic-ray-induced reactions, formation of \molH{} on dust grains, and freeze out/thermal desorption of CO and \water{} onto/from dust grains.

Note that we exclude S from our model as its abundance and main carriers in protoplanetary discs are somewhat debated. Although previous models used the ISM measurements of \citet{Savage_1996} to determine gas-phase abundances - in which S was not measured to be depleted with respect to solar - S does appear to become progressively depleted as material becomes increasingly dense during the molecular cloud and core phases of star formation \citep{Hily-Blant_2022,Fuente_2023}. Similarly, protoplanetary discs exhibit gas-phase S depletion \citep{Kama_2019} of around 90 percent, implying a highly abundant refractory reservoir; meteoritic studies show abundant FeS minerals in chondrites \citep{Kallemeyn_1989}, which may therefore be a likely candidate. As a consequence, previous models of wind emission overestimate how bright and frequently-detected S emission lines should be \citep{Pascucci_2020} and how much S can contribute to cooling.

\subsection{Atomic cooling}
\label{sec:methods_atomic}
Atomic line cooling happens via collisional excitation of an atom or ion to an electronically excited state, followed by radiative de-excitation.
To find the total cooling emission, we therefore need to compute the population of each atomic level. To do so, we assume equilibrium between collisional and radiative transitions and solve the linear system formed by the collisional (de-)excitation rates and the spontaneous emission rates. In particular, the relative abundance of the $i$th level $x_i$ is determined by
\begin{equation}\label{eqn:linear_system_cooling}
-x_i\sum_\ell\left( n_\ell\sum_{i\neq j}k_{ij}^\ell\right) - x_i\sum_{j<i} A_{ij} + \sum_\ell\left( n_\ell\sum_{i\neq j}k_{ij}^\ell x_j\right) + \sum_{j>i} A_{ji} x_j = 0\,,
\end{equation}
where $n_\ell$ is the abundance of the $\ell$th collider ($\ell \in \{\mathrm{e^-,\,H,\,H^+,\,ortho-H_2,\,para-H_2}\}$), $k_{ij}^\ell$ is the temperature-dependent collisional rate coefficient that takes into account the transition between the $i$th and $j$th levels when (de-)excited by the $\ell$th collisional partner, and $A_{ij}$ is the Einstein coefficient of the $i\to j$ radiative de-excitation.

By solving the linear system using Eq.\,(\ref{eqn:linear_system_cooling}) for each level and replacing the last level equation with the mass conservation (i.e. $\sum_i x_i = 1$), it is possible to compute the level population $x_i$. This system of equations can be solved with a matrix approach as discussed in \ref{sec:methods_atomicSolver}.

After $x_i$ are computed, the cooling emission of each transition becomes ${\Lambda_{ij}= n\, A_{ij}\, x_i\,\Delta E_{ij}}$, where $n$ is the total abundance of the specific atomic coolant, and $\Delta E_{ij}$ is the energy difference between the $i$th and $j$th levels.
All of the transitions included here (with the exception of Lyman $\alpha$, discussed below), are forbidden transitions, to which the wind is generally optically thin and thus we assume an escape probability of unity. To confirm this, we checked the line-centre cross-section of the lines assuming a thermal broadening $v_{rm th}\sim1\,\mathrm{km\,s^{-1}}$ according to $\sigma = \frac{\lambda^3 A_{21}}{8\sqrt{2}\pi^{3/2}v_{\rm th}}$ \citep[][Equation 4.45]{Osterbrock}. For the shorter wavelength optical lines this is typically $\lesssim 10^{-21}\,\mathrm{cm^2}$, while those of the FIR lines are somewhat higher at $\lesssim 10^{-17}\,\mathrm{cm^2}$ (the highest being \ion{O}{I} $145\,\mathrm{\mu m}$ with $8.6 \times 10^{-18}\,\mathrm{cm^2}$. However, given the atomic abundances of $\sim 10^{-4}$, the cross-sections per hydrogen atom are $\lesssim 10^{-21}\,\mathrm{cm^2}$. Therefore, given that soft X-rays typically penetrate to $\sim 10^{21}\,\mathrm{cm^{-2}}$, we expect the lines to be at most marginally optically thick in the heated wind gas.
The total cooling is the sum of all the allowed (or relevant) atomic transitions.

In Table \ref{tab:atomicCool}, we summarise the number of levels used for each of our atomic coolants (O, O\textsuperscript{+}, C, C\textsuperscript{+}) and the provenance of the data for the collisional (de-)excitation strengths/rates. As detailed above, we include $\mathrm{e^-,\,H,\,H^+,\,ortho-H_2,\,and\,para-H_2}$ as colliders wherever data is available. This ensures that we capture collisional excitement and cooling as far as possible in ionised, neutral atomic and molecular environments.
In most cases, suitable numerical fits to the appropriate rates calculated at different temperatures were available in \citet[][Appendix F]{Draine_2011}. For excitation of O and C by H\textsuperscript{+} we used piece-wise fits from \citet{Glover_2007}.
For collisional (de-)excitation of O by H, ortho-\molH{} and para-\molH{}, we use the latest calculations by \citet{Lique_2018}\footnote{Note that the values quoted in the paper have been corrected \citep{Wolfire_2022}, and are available in the Leiden Atomic and Molecular Database \citep[LAMDA,][]{LAMDA_2020} at \url{https://home.strw.leidenuniv.nl/~moldata/datafiles/oatom@lique.dat}.} to which we perform our own fits, presented in Appendix \ref{appendix:Lique}.
Level energies, degeneracies, and Einstein coefficients of spontaneous emission $A_{ij}$, were taken from the National Institute of Standards and Technology database \citep{NIST}.

\begin{table*}
    \caption{References for the atomic collisional rate data used by our version of \textsc{prizmo}.}
    \label{tab:atomicCool}
    \centering
    \setlength\tabcolsep{3pt}
    \begin{tabular}{llllll}
         \hline\hline
         Species & $n_{\rm levels}$ & \multicolumn{4}{c}{Source of Collisional Rate Data by Collider} \\
         & & e\textsuperscript{-} & H & H\textsuperscript{+} & ortho-\molH{} \& para-\molH{} \\
         \hline
         O & 5 &
         \term{3}{P}{$J_j$}-\term{3}{P}{$J_i$}: \citet{Bell_1998}; &
         \term{3}{P}{$J_j$}-\term{3}{P}{$J_i$}: \citet{Lique_2018}; & \term{3}{P}{$J_j$}-\term{3}{P}{$J_i$}:  & \term{3}{P}{$J_j$}-\term{3}{P}{$J_i$}:  \\
         & & otherwise \citet{Pequignot_1990} & \term{1}{D}{2}-\term{3}{P}{$J_i$}: \citet{Krems_2006} & \citet{Glover_2007} & \citet{Lique_2018} \\
         O\textsuperscript{+} & 5 &
         \citet{Tayal_2007} & ... & ... & ... \\ 
         C & 3 &
         \citet{Glover_2007} & \citet{Abrahamsson_2007} & \citet{Glover_2007} & \citet{Schroder_1991} \\
         C\textsuperscript{+} & 2 &
         \citet{Tayal_2008} & \citet{Barinovs_2005} & ... & \citet{Flower_1977} \\
         \hline
    \end{tabular}
\end{table*}

At the temperature range of interest $\lesssim10^4~\mathrm{K}$, we expect emission from lines with excitation temperatures up to a few times $\lesssim10^4~\mathrm{K}$, that is, with wavelengths into the optical. If these optical coolants are missing, wind temperatures may be overestimated \citep{Sellek_2022}. Consequently, we include all of the fine-structure terms in the ground configuration of O and O\textsuperscript{+} (resulting in 5-level systems) such that in both neutral and ionised environments, we can have optical line cooling.
However, we keep C and C\textsuperscript{+} to a 3- and 2-level system, respectively, due to the limited availability of collision strengths and the increased computational overhead of calculating the level population for higher-level systems; O has typically been found to be a more important coolant than C.

By comparing to a list of the most important coolants in the wind region in the fiducial \textsc{mocassin} model of \citet{Sellek_2022}, we confirmed that modulo the above caveats, our approach captures most of the lines that contribute to cooling at more than the percentage level, and that beyond this the model is limited by the absence of additional coolants (e.g. S, Si, Fe) rather than the number of levels.

Lyman $\alpha$ cooling is included according to \citep{Spitzer_1978} and assumed to be effectively thin as at the densities of interest, any reabsorbed photons are likely to be quickly re-emitted or reprocessed to optically thin wavelengths by dust \citep{Sellek_2022}.

\subsection{Spectra}
\begin{figure}
    \centering
    \includegraphics[width=\linewidth]{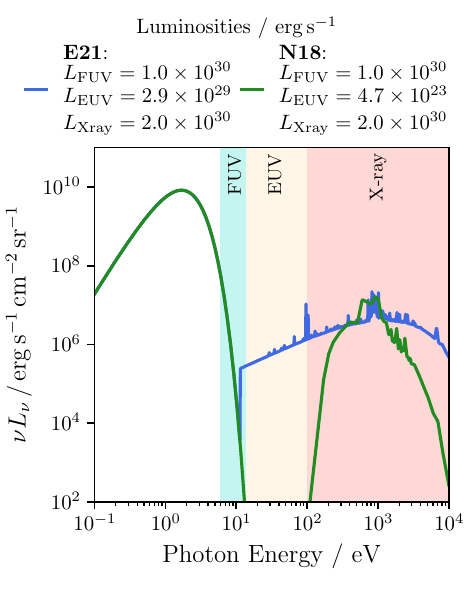}
    \caption{Comparison of the two spectra \citep[E21:][blue]{Ercolano_2021}; \citep[N18:][green]{Nakatani_2018b} investigated in this work with different regions of the spectrum highlighted and the respective luminosities given in the legend.}
    \label{fig:spectra}
\end{figure}

The choice of the irradiating spectrum can be highly important for the outcome of photoevaporation simulations \citep{Sellek_2022}. To enable comparison with the most recent generation of X-ray photoevaporation models, our fiducial spectrum is the ``Spec30'' X-ray spectrum of \citet{Ercolano_2021}. We normalise this spectrum to a typical $L_X = 2.04\times10^{30}\,\mathrm{erg\,s^{-1}}$ (as measured between $0.5-5\,\mathrm{keV}$) for a solar-mass star as per \citet{Picogna_2021}. To this, we also add a 5000 K blackbody assuming a stellar radius of $1\,{\rm M}_{\sun}$; this is to ensure reasonable dust temperatures through heating by optical/UV photons. The spectrum is shown in Fig.~\ref{fig:spectra}.

Our fiducial spectrum is used throughout our benchmarks, our main simulation and our explorations of cooling and microphysics.
In addition, we run a single simulation (see Sect. \ref{sec:spectrum}) with the X-ray spectrum used by \citet{Nakatani_2018b,Komaki_2021} - which is based on a model of the observed X-ray spectrum of TW Hya \citep{Nomura_2007} - which we normalise and add to the blackbody in the same way as the fiducial spectrum.

\subsection{Dust}
\label{sec:dust}
The dust number density $n_{\rm d}$ was assumed to follow a grain size distribution of $\varphi(a) = \frac{\dd n_{\rm d}(a)}{\dd a}\propto a^{-3.5}$ power law between $5\times10^{-7}$ and $2.5\times10^{-5}$ cm  \citep{Mathis_1977}, with bulk density of $3\,\mathrm{g\,cm^{-3}}$ \citep[e.g., see][]{Zhukovska_2008,Grassi_2017}. The optical properties are calculated with the Mie scattering theory using the dielectric functions of \citet{Draine_2003a,Draine_2003b} which assume a mix of a graphitic carbonaceous component and amorphous astronomical silicates.

These properties are all suggestive of ISM-like grains which have not undergone grain growth. As such, all these grains would be expected to be entrained by a wind \citep[e.g.][]{Hutchison_2016a,Franz_2020,Booth_2021} so we assume a uniform ISM dust-to-gas mass ratio of $10^{-2}$ everywhere. However, note that grain growth is expected to occur within $10^5\,\mathrm{yr}$ of disc evolution and thereafter, maybe as little as $\lesssim10\%$ of the dust mass can enter the wind \citep{Franz_2022a}. Our results, therefore, likely overestimate the abundance of dust in the wind and hence provide an upper limit on its effects on the thermal structure. While we test our sensitivity to this in Sect. \ref{sec:benchmarks_Xray} and discuss what a more reasonable value might be in light of our results in Sect. \ref{sec:dustentrainment}, a more consistent approach with variable dust entrainment is beyond the scope of this exploratory study.

\section{Methods: hydrodynamics}
\label{sec:methods_hydro}
\subsection{Overall workflow}
Our hydrodynamic simulations are conducted in a three-stage process using \textsc{pluto} \citep{Mignone_2007}.

We start with an initial disc model \citep[originally presented in][]{Picogna_2021} generated using the D’Alessio Irradiated Accretion Disc (DIAD) models \citep{DAlessio_1998,DAlessio_1999,DAlessio_2001,DAlessio_2006} for a gas disc mass of $0.045\,{\rm M}_{\sun}$ (extending to $400$ au) orbiting a K6 star (with $M_\star=1\,{\rm M}_{\sun}$, $L_\star=2.335\,{\rm L}_{\sun}$, $R_\star=2.615\,{\rm R}_{\sun}$, $T_\star=4278\,\mathrm{K}$).
The dust is assumed to be well-mixed, with a maximum grain size in the atmosphere of $2.5\times10^{-5}$ cm and a minimum grain-size of $5\times10^{-7}$ cm (the same as we assume in \textsc{prizmo}). 
We use this to calculate initial conditions for the wind by evolving the hydrodynamics for 50 orbits at 10\,au, with the wind temperatures calculating temperatures using the ``Spec30'' $\xi-T$ relationship of \citet{Ercolano_2021}. Aside from details of the extent/resolution of the numerical grid, this is effectively identical to the $1\,{\rm M}_\sun$ case from \citet{Picogna_2021} and so we refer to this as our \citet{Picogna_2021} analogue simulation.

We use the \citet{Picogna_2021} analogue simulation to provide a density input to \textsc{pluto+prizmo} that already has a wind such that the initial conditions are likely to be closer to the solution; this results in a smoother and more rapid approach towards the steady state without producing transient features due to the sudden expansion of the heated disc material into the wind region.

Secondly, we then run \textsc{pluto+prizmo} with all the hydrodynamic processes disabled in order to perform a ``burn-in'' in which only the chemical network and temperature are evolved for 0.1 Myr. This allows us to reach an equilibrium where these quantities are consistent with the density profile. This also allows us to benchmark the temperatures calculated by \textsc{prizmo} against the temperatures prescribed from the $\xi-T$ relationships and identify areas of major difference (see Sect. \ref{sec:benchmarks_Xray}).

Finally, the hydrodynamic processes are re-enabled.
The simulation is first run for 50 orbits at 10\,au (approximately one gravitational radius for $10^4\,\mathrm{K}$ gas).
Since the structure and composition of the wind settle from the inside out, we are then able to freeze the inner half of the grid ($r < 17.3 \, \mathrm{au}$) in order to speed up the calculation during a further 150 orbits.

\subsection{Hydrodynamics-thermochemistry interface}
We perform several necessary modifications to PLUTO to couple it with \textsc{prizmo}\footnote{This version of the code is available from GitHub at: \url{https://github.com/AndrewSellek/PLUTO_PRIZMO/tree/chemistry}, commit \texttt{8e525b9}}.

The hydrodynamics and thermochemistry are solved alternately within \textsc{pluto}'s Strang splitting scheme. On even-numbered steps, hydrodynamics, including advection of species, first takes place over the hydrodynamic time step $\delta t$; assuming no errors are encountered (see below), \textsc{prizmo} is then called as part of the ``SourceStep''. On odd-numbered steps, the order of the two processes is reversed with thermochemistry taking place first followed by the hydrodyanmics.

The mass fraction of each species in the network is tracked during the hydrodynamics using \textsc{pluto}'s tracer variables. By default, \textsc{pluto} does not know that these should - like the density - be non-negative and therefore in principle \textsc{prizmo} can receive negative abundances leading to a fatal error. To avoid this, we added checks to \textsc{pluto}'s ``ConsToPrim'' function where similar checks are carried out on the energy, pressure and density. If the tracer mass-fraction $\epsilon_i>-1\times10^{-6}$ and the tracer density $\rho_i=\epsilon_i\rho>-\rho_{\rm floor}$ the mass-fraction is simply set to 0, otherwise the cell is flagged for recalculation, first using different solvers (\textsc{pluto}'s default FAILSAFE method), or, if that fails, by reducing the timestep by a factor 10 until significant negative values are no longer produced. If no solution is found after ten retries, we consider that the solution does not converge so the code returns an error and aborts; once in our fiducial run we required four retries otherwise no more than two were ever needed.

Within the thermochemical step, the first task is to update the radiation field. For this we conduct radial tracing, considering both geometric dilution and the attenuation:
\begin{equation}
    j_{\nu}^{i+1} = j_{\nu}^{i+1} e^{-\tau_\nu} \frac{r_i^2}{r_{i+1}^2}.
\end{equation}
The attenuation $e^{-\tau_\nu}$ calculated by \textsc{prizmo} includes absorption of radiation by the dust as well as all relevant gas-phase species.
Note that by conducting radial ray tracing we neglect the scattering of radiation which can be important for UV photons but is negligible for the X-ray photons that ultimately drive the wind we see. Subsequently, \textsc{prizmo} then advances the chemical network in each cell of the simulation for a total time $\delta t$. In so doing, the gas temperature is evolved along with the number densities of each species; the updated values are passed back to \textsc{pluto} in the form of the gas pressure and the tracer variables.

However, within the underlying disc, the dust radiative transfer is an inherently diffusive problem.
Since we only conduct 1D radiative transfer with no scattering of radiation - rather than adopting an expensive Monte Carlo \citep[e.g.]{MCFOST,RADMC3D} or Flux-Limited Diffusion \citep{Levermore_1981,Kuiper_2010} approach - \textsc{prizmo} alone significantly underpredicts dust temperatures below the optical photosphere. This is a problem since, due to the tight dust-gas coupling expected here, the gas temperature would be similarly low and hence the disc would lose a lot of vertical pressure support and shrink. We thus follow the typical approach, which \citep{Owen_2010,Wang_2017} is to pin the temperature structure to a pre-calculated dust temperature structure \cite[e.g.,][]{Chiang_1997,DAlessio_2001} beyond a certain column density. Therefore at high column densities we smoothly interpolate between the \textsc{prizmo} temperatures $T_{\rm PRIZMO}$ and the initial conditions \citep[which given the equivalent pinning done in the $\xi-T$ step will be the][temperature, $T_{\rm DIAD}$]{DAlessio_1998,DAlessio_1999,DAlessio_2001,DAlessio_2006} according to the following equation (where $N_{23} = N_{\rm tot}/10^{23}\,\mathrm{cm^{-2}}$):
\begin{equation}
    T_{\rm gas} = 
    \begin{cases} 
      T_{\rm PRIZMO} & N_{\rm tot}\,/\,\mathrm{cm^{-2}} \leq 10^{21} \\
      \frac{T_{\rm PRIZMO}+N_{23}^2T_{\rm DIAD}}{1+N_{23}^2} & 10^{21} \leq N_{\rm tot}\,/\,\mathrm{cm^{-2}} \leq 10^{25} \\
      T_{\rm DIAD}   & 10^{25} \leq N_{\rm tot}\,/\,\mathrm{cm^{-2}}\,.\\
   \end{cases}
   \label{eq:Tweighting}
\end{equation}
We show and discuss an example of how this affects the dust/gas temperature in Sect. \ref{sec:results}.

This method also significantly benefits the computational cost by speeding up the calculation in high-density cells, since the temperature and chemistry are no longer coupled.

\subsection{Hydrodynamics setup}
We conduct the full hydrodynamics+thermochemistry simulations on a 2D spherical grid with 400 cells in the radial direction extending from $0.75-400\,\mathrm{au}$ and 70 in the latitudinal direction between the midplane and polar axis (spaced to maximise the resolution in the wind-launching region).
For the burn-in with thermochemistry only (and \citet{Picogna_2021} analogue), we extend the grid inwards to $r=0.25\,\mathrm{au}$ using an extra 70 cells. This is to ensure a steady solution at the inner radial boundary, as testing showed that if the inner cells received completely unattenuated radiation, the results were quite unstable as the disc would develop a very thin, hot region in the inner row of cells that should not be present (assuming a full disc has no cavity). The region $0.25-0.75\,\mathrm{au}$ therefore acts to attenuate the radiation reaching the disc and the subsequent hydrodynamic solution is obtained using the radiation field measured in the burn-in runs at $0.75\,\mathrm{au}$.

In the latitudinal direction we used \textsc{pluto}'s ``polaraxis'' and ``eqtsymmetric'' boundary conditions at the rotation axis and disc midplane.
At the outer radial boundary, we used outflowing boundary conditions, while at the inner radial boundary we adopted user-defined conditions to ensure stability; these are identical to outflowing conditions except that in order to prevent material entering the grid from radii where we do not expect any wind, the radial component of velocity is set in the manner of reflective boundary conditions if it has a positive sign.

Finally, we used the following \textsc{pluto} settings: linear reconstruction, RK2 timestepping, the MINMOD\_LIM limiter and the HLLC solver. Moreover, since early tests with our code were susceptible to negative pressures (a common problem which can result from the fact that the usual equations of hydrodynamics aim to conserve total and kinetic energy and obtain the thermal pressure as the difference, which can become negative due to discretisation errors) we used \textsc{pluto}'s option to instead conserve entropy, which conserves thermal energy/pressure more directly. We found that using this option adequately avoided these errors. 

\section{Static benchmarks of \textsc{prizmo}}
\label{sec:benchmarks}
To validate the working of \textsc{prizmo} we first perform two benchmarks in the limits corresponding to the extremes of literature photoevaporation models: a low-density/high ionisation parameter EUV model from \citep{Wang_2017} and a higher-density X-ray model from our \citet{Picogna_2021} analogue.

\subsection{High ionisation parameter regime}
\label{sec:benchmarks_EUV}
\citet{Sellek_2022} previously benchmarked the temperature structure of the EUV wind of \citet{Wang_2017} against \textsc{mocassin}. This work demonstrated that the wind temperatures were strongly overestimated by \citet{Wang_2017} compared to \textsc{mocassin} due to a lack of effective optical lines available for cooling in their model: this could affect the radii at which a wind can launch, as well as the dominant role for adiabatic cooling that they concluded, and the chemical composition of the wind. On the other hand, \textsc{mocassin} found higher wind temperature just below the wind base, likely as a result of it not including molecular cooling.

\begin{figure*}
    \centering
    \includegraphics[width=\linewidth]{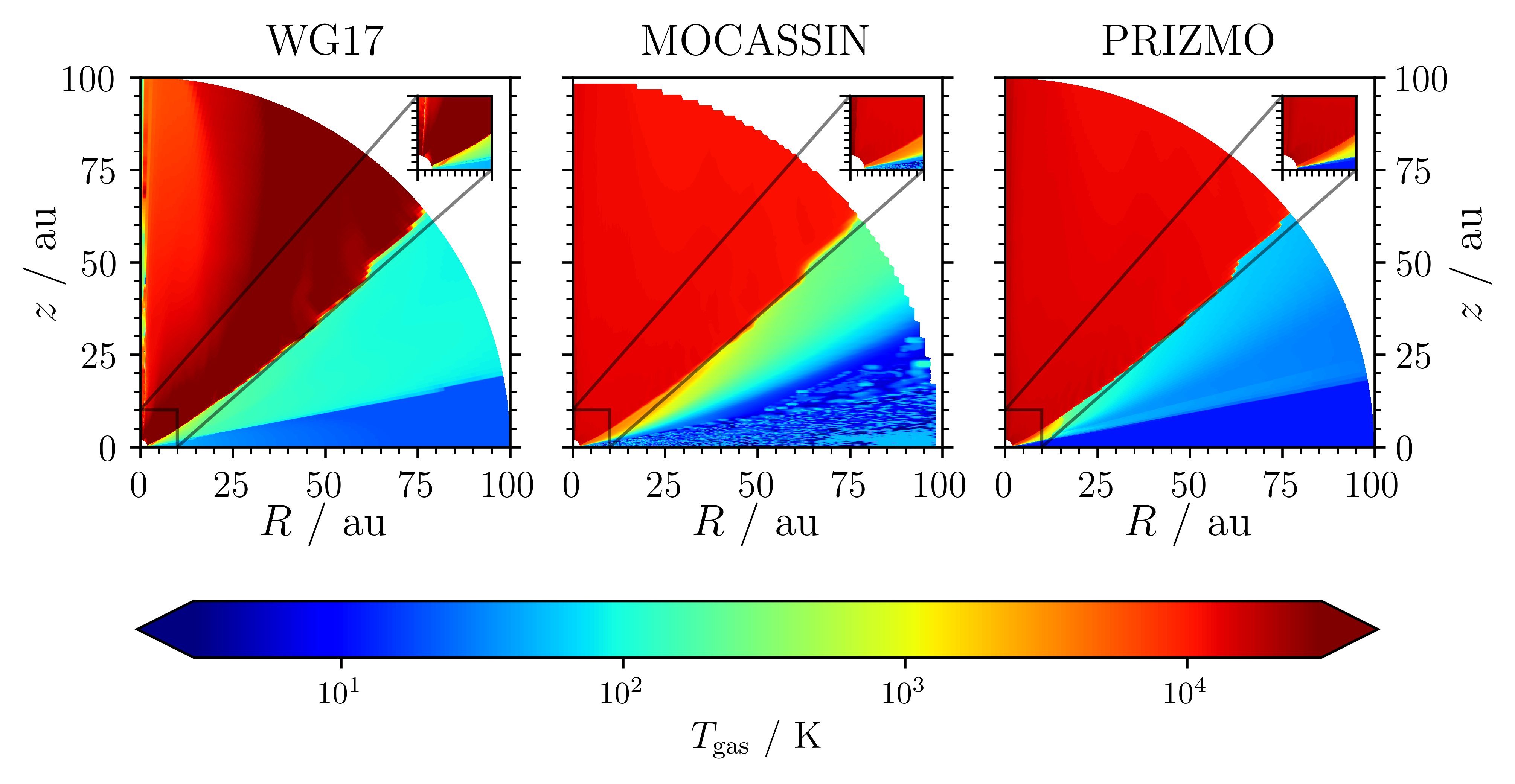}
    \caption{Comparison of the gas temperatures predicted by (L-R) \citet{Wang_2017}, \textsc{mocassin} \citep{Sellek_2022} and \textsc{prizmo} (this work) for the EUV-wind density model of \citet{Wang_2017}. Inset panels show the inner 10 au.}
    \label{fig:benchmark_WG17}
\end{figure*}

In Fig.~\ref{fig:benchmark_WG17}, we compare the gas temperatures calculated by \textsc{prizmo} to the original temperatures of \citet{Wang_2017} and those calculated by \citet{Sellek_2022}. Since \textsc{prizmo} is designed to be the most comprehensive of the three codes - covering atomic cooling at low and high temperatures in neutral and ionised regions and also molecular cooling in warm dense gas - it is no surprise that everywhere it generally agrees with the cooler of the other two codes.
Thus, in the atomic wind, optical lines help \textsc{prizmo} successfully avoids the overly high $2-3\times10^4\,\mathrm{K}$ temperatures predicted by \citet{Wang_2017} and thus agree with \textsc{mocassin}. Conversely, just below the wind base, molecular cooling in \textsc{prizmo} shrinks the large warm region at 10s au predicted by \textsc{mocassin}, instead limiting it to the inner $\sim25$ au as seen by \citet{Wang_2017}.

Note that deeper into the disc where the dust and gas temperatures are coupled, \textsc{prizmo} and \citet{Wang_2017} continue to find even lower temperatures than \textsc{mocassin}, which cannot result from the additional cooling. This is since, unlike \textsc{mocassin}, \textsc{prizmo} and \citet{Wang_2017} do not calculate the diffuse radiation field that heats the dust, and hence underestimate the dust (and coupled gas) temperatures below the optical photosphere. Consequently, to avoid this problem, the temperatures \citet{Wang_2017} report - and which we plot in Fig.~\ref{fig:benchmark_WG17} - are set to be the maximum of that predicted from radial ray tracing alone and the temperature profile from \citet{Chiang_1997}, which reduces the extent of the underestimate. Therefore, \textsc{prizmo} shows the coolest temperatures; this would affect the hydrostatic support of the disc and hence when using the temperatures for hydrodynamics we similarly have to pin the temperature as described by Eq.~\ref{eq:Tweighting}.

Overall, we conclude that \textsc{prizmo} performs well in the two extremes of the low-density winds with high ionisation parameter and in the dense underlying disc.

\subsection{Lower ionisation parameter regime}
\label{sec:benchmarks_Xray}
\begin{figure*}
    \centering
    \includegraphics[width=\linewidth]{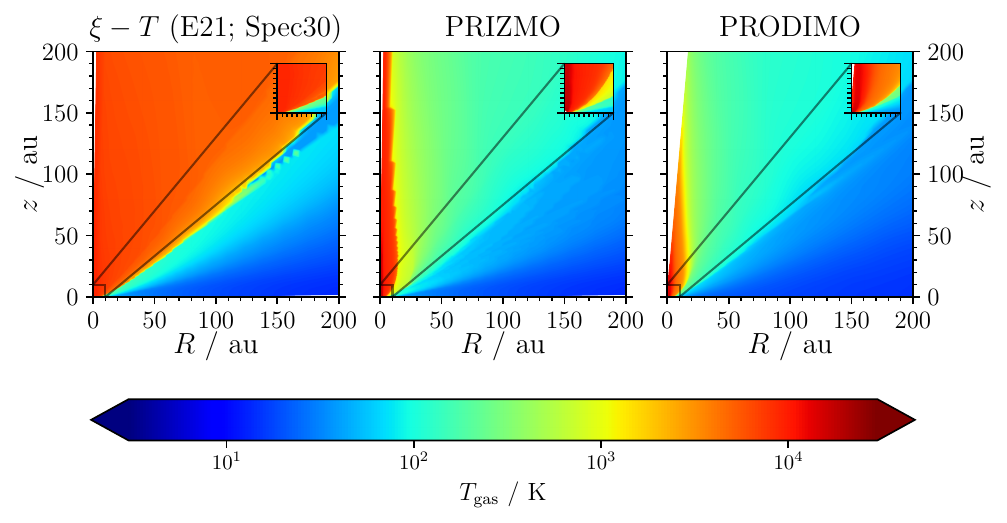}
    \caption{Comparison of the gas temperatures predicted by (L-R) the $\xi-T$ relationship from \citet{Ercolano_2021} \citep[c.f.][]{Picogna_2021}, \textsc{prizmo} (this work) and \textsc{ProDiMo} for the \citet{Picogna_2021} analogue model. Inset panels show the inner 10 au.}
    \label{fig:benchmark_P21}
\end{figure*}

Previous generations of X-ray photoevaporation models have generally relied on the ionisation parameter to control the relationship through a prescribed relationship. The temperatures after 50 orbits of our \citet{Picogna_2021} analogue, which are calculated in this way, are shown in the leftmost panel of Fig.~\ref{fig:benchmark_P21} - they show the typical appearance of an X-ray--driven wind: a nearly isothermal temperature of several 1000 K.

In the central panel of Fig.~\ref{fig:benchmark_P21}, we compare these to the temperatures calculated by post-processing this model using \textsc{prizmo} during the ``burn-in'' step. These are vastly different, showing generally cooler temperatures and a strong radial dependence. In the right-hand panel, we confirm that this is not a particular issue with \textsc{prizmo} by showing that post-processing with another thermochemical code \textsc{ProDiMo} \citep[PROtoplanetary DIsk MOdel\footnote{\url{https://prodimo.iwf.oeaw.ac.at} \mbox{rev.: 4fee3902 2023/06/19}},][]{Woitke_2009,Kamp_2010,Thi_2011,Woitke_2016,Rab_2018} shows the same qualitative features. 
We note that the \textsc{ProDiMo} model uses a larger chemical network (100 species, see \citealt{Kamp_2017} for details) and includes heavy elements such as Fe or S, and more complete atomic line lists. Hence, the temperatures from \textsc{ProDiMo} tend to be a bit cooler. Nevertheless, the overall agreement of PRIZMO with \textsc{ProDiMo} is very good, indicating that the choices for the reduced chemical network and simplified heating/cooling are well justified. 
%(although tends to be a little cooler as their atomic line list is more complete).

\begin{figure*}
    \centering
    \includegraphics[width=\linewidth]{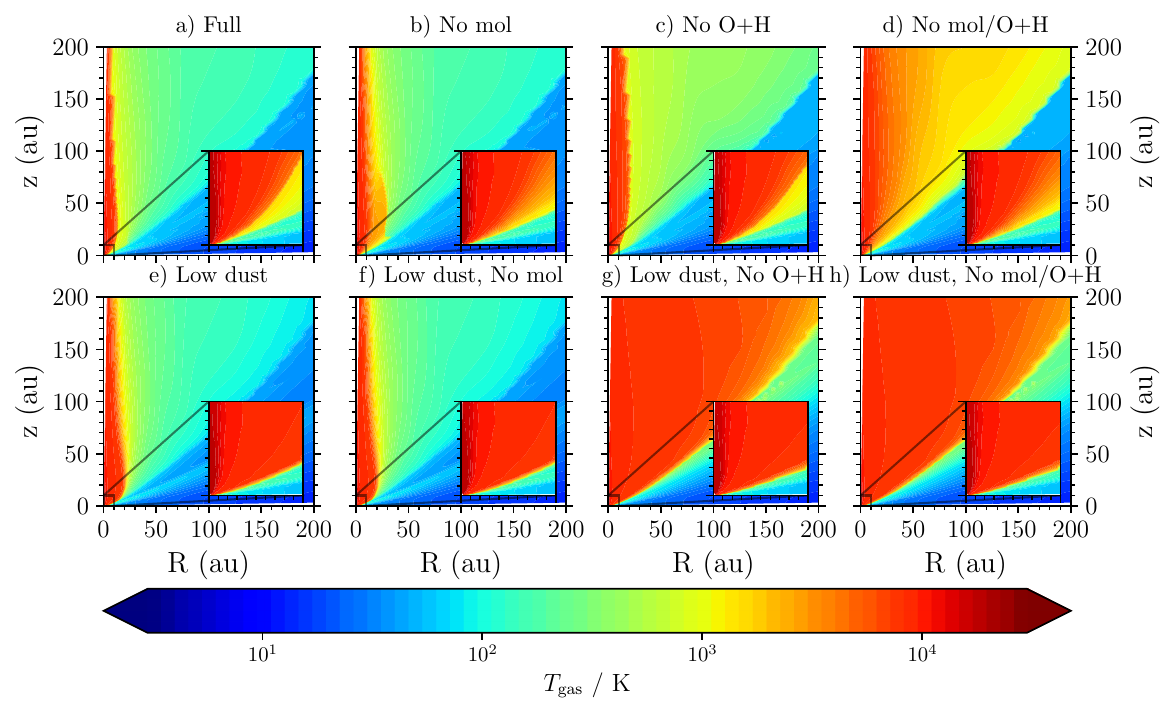}
    \caption{Comparison of the temperatures predicted \textsc{prizmo} when various combinations of cooling processes are switched off or reduced: "No Mol" means that molecular cooling due to \molH{} and CO are not included; "No O+H" means that collisional excitation of O by H is not included; "Low dust" means that the dust-to-gas mass ratio was lowered to $10^{-10}$. Inset panels show the inner 10 au.}
    \label{fig:microphysics}
\end{figure*}

The cooler temperatures in this regime suggest that there are cooling channels included in \textsc{prizmo} and \textsc{ProDiMo} but not in \textsc{mocassin}, hence we now experiment with turning off different (combinations of) processes in \textsc{prizmo} to see what drives the difference.
In this, we must keep in mind two questions with potentially separate answers: a) which cooling destabilises the temperatures from the $10^4\,\mathrm{K}$ temperatures predicted by the $\xi-T$ relationship? and b) and which ends up dominating the cooling in the final equilibrium?

The most obvious candidate to make a difference might be molecular cooling. However, since \molH{} and CO are readily photodissociated (and, unlike in later sections, are not replenished here by advection as this is a static benchmark), their abundances are not high enough to dominate the cooling. Thus, in panel b) of Fig.~\ref{fig:microphysics}, turning off the molecular cooling does not raise the temperatures significantly save for in a limited region around $20-30\,\mathrm{au}$. Moreover, especially since thermal dissociation makes molecular survival at high temperatures even less likely, molecular cooling cannot be the process that destabilises the hot wind temperatures.

Another difference is that the version of \textsc{mocassin} that was used in the works described above only includes collisional excitement by electrons, whereas in this work, we include electrons, H, H\textsuperscript{+}, and both ortho-/para-\molH{}. Oxygen is generally found to be a bright, dominant coolant and H will be the most abundant collider in a largely neutral atomic wind; indeed \citet{Ercolano_2016} found that these collisions were necessary to match observed [\ion{O}{I}] line luminosities. Therefore, in panel c) of Fig.~\ref{fig:microphysics}, we consider the effect of switching off the collisions between O and H. We see that this has something more of an effect in raising the temperature on scales of 10s-100s au, though still note much above 1000 K.
This suggests that this is the process dominating the cooling in the final equilibrium, but alone is not destabilising the hottest temperatures.
Switching off both molecular cooling and collisions between O and H is somewhat more powerful than either alone (Fig.~\ref{fig:microphysics}, panel d). This is in line with the fact that \molH{} cooling is highly sensitive to the temperature at 100s K and will be boosted by the small rises; moreover, the decrease in the excitation of the oxygen lines due to fewer colliders limits the role of atomic cooling. The overall result is that \molH{} cooling is therefore what dominates at the ``new'' equilibrium shown in panel c), and can cool the wind quite considerably but not quite so much as the collisional excitation of O by neutral H (and still cannot be destabilising the hottest temperatures). However, even without both processes, the temperatures are still lower than predicted from the $\xi-T$ relation (Fig.~\ref{fig:benchmark_P21}).

Furthermore, our model includes a large amount of small dust; in the wind $T_{\rm dust}<T_{\rm gas}$ and hence it can have a cooling effect. To see the effect of this, we reduce the dust-to-gas mass ratio to a negligible $10^{-10}$. This has relatively little effect on its own (Fig.~\ref{fig:microphysics}, panel e) or in combination with switching off molecular cooling (Fig.~\ref{fig:microphysics}, panel f). This further underscores the fact that collisional excitation of O by neutral H is dominating the cooling at the equilibrium achieved by the full \textsc{prizmo} model. Moreover, as argued above for O+H collisions, thermal accommodation with dust cannot be the sole process destabilising the hot temperatures. However, when depleted dust is used in combination with switching off collisions between O and H (Fig.~\ref{fig:microphysics}, panel g), we find approximately isothermal several 1000 K temperatures comparable to those predicted by the \citet{Ercolano_2021} $\xi-T$ relation (very little additional difference is made by also removing molecular cooling; Fig.~\ref{fig:microphysics}, panel h). This confirms that both thermal accommodation with dust and collisional excitation of O by neutral H can destabilise the hot temperatures and both must be removed to prevent this, while cooling by \molH{} does not contribute to this picture and only dominates the cooling under conditions where it is cool enough to be abundant enough but warm enough to excite its emission lines.

To summarise,
\begin{itemize}
    \item either thermal accommodation with dust or collisional excitation of O by neutral H can destabilise the hot $10^4\,\mathrm{K}$ temperatures and both processes must be removed from \textsc{prizmo} to recover them; \molH{} cooling is irrelevant here as it would not survive at equilibrium at such high temperatures.
    \item Collisional excitation of O by neutral H dominates the cooling at the thermal equilibrium reached in the full model; thus removing the other processes alone has little effect on the temperatures.
    \item Once the temperatures are cool enough for some \molH{} survival, it is the next most important coolant after collisional excitation of O by neutral H and prevents the temperatures rising too significantly if only O+H collisions are removed.
\end{itemize}

Therefore, in conclusion, the cooler temperatures result for two combined reasons, one more fundamental, one more related to the modelling choices we make: while the dust abundance and grain size distribution could reasonably be such that dust is not a significant coolant (in this case panel e) would be most realistic), we cannot escape the fact that \textsc{mocassin} underestimated the atomic cooling in largely neutral media (where H is much more abundant than electrons) due to its omission of O+H collisions. O+H collisions are sufficient to lower the temperature in the intermediate density/ionisation parameter conditions seen at 10s au in previous X-ray photoevaporation models, and dominate the cooling in these static models so long as the process is included.
Conversely, wherever the degree of ionisation is high, such as inside $10-20\,\mathrm{au}$ where the X-ray flux is large (or in the lower-density EUV photoevaporation scenario; Sect. \ref{sec:benchmarks_EUV}), the relative contribution of O+H collisions to the cooling rate is small, and the optical lines (including the Lyman lines of H) lead to the usual $10^4\,\mathrm{K}$ thermostat.
The result is that lower densities would be needed to achieve the same temperatures seen in previous photoevaporation models, this could be visualised as translating the $\xi-T$ relation to higher ionisation parameters (see Sect. \ref{sec:explainxiT}).

\section{Results}
\label{sec:results}
\subsection{Structure of the wind}

\begin{figure*}
    \centering
    \includegraphics[width=\linewidth]{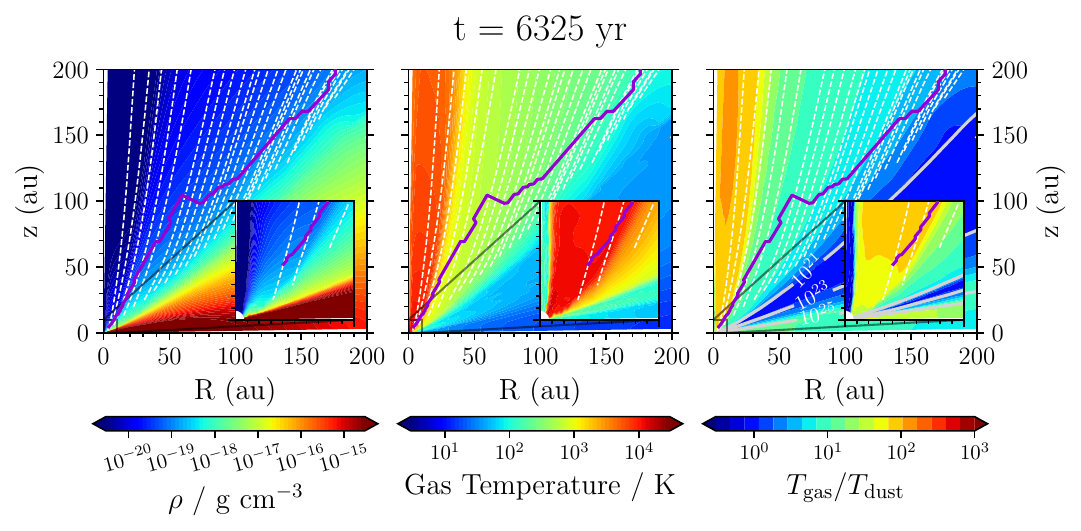}
    \caption{The gas density (left), gas temperature (centre) and ratio of gas and dust temperatures (right) after 200 orbits of the fiducial simulation. The latter is defined as $\min(T_{\rm gas, PRIZMO},T_{\rm DIAD})/T_{\rm dust, PRIZMO}$.
    The solid purple line is the sonic surface, while the dashed lines show streamlines at regular intervals.
    The grey lines in the right-hand panel show total column densities of $N=10^{21},10^{23},10^{25}\,\mathrm{cm^{-2}}$.}
    \label{fig:rho_Tgas_deltaT}
\end{figure*}

\begin{figure*}
    \centering
    \includegraphics[width=\linewidth]{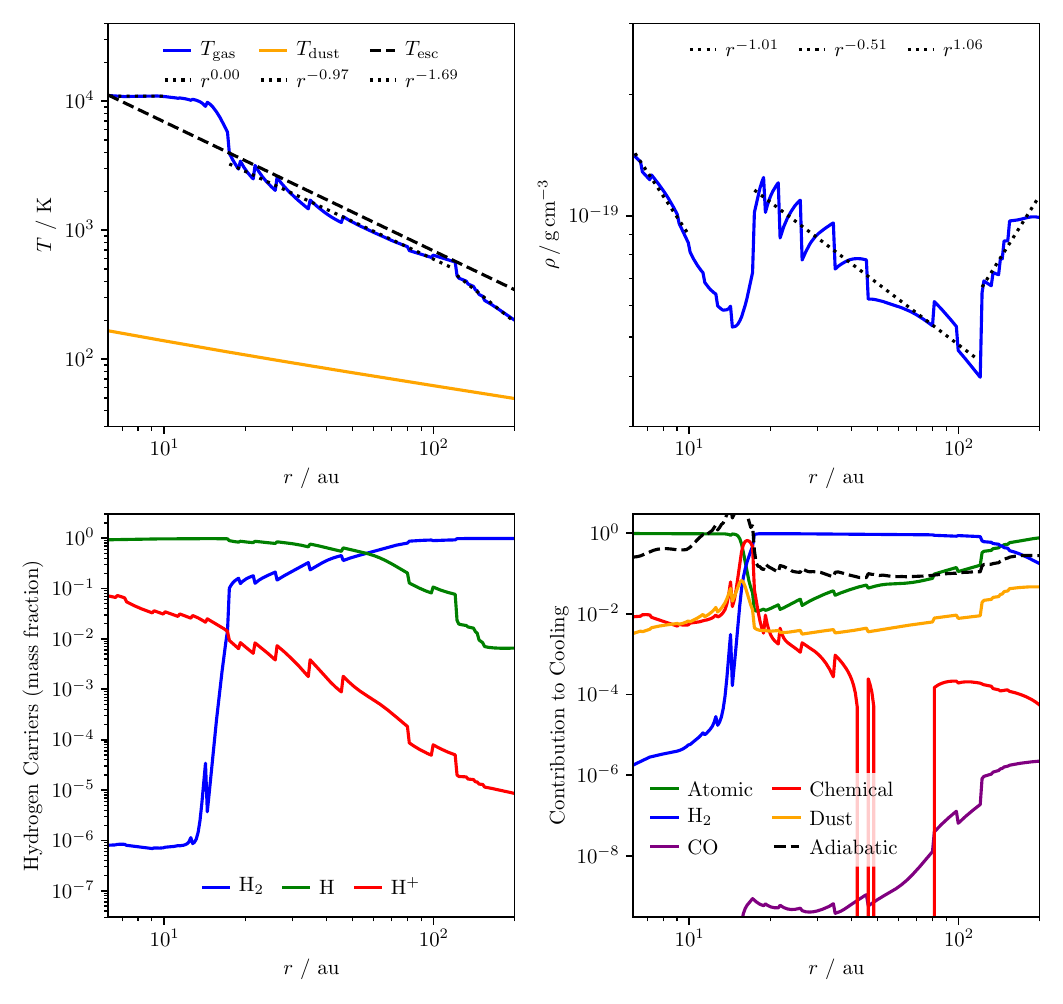}
    \caption{Quantities extracted along the sonic surface. We only show $r>6\,\mathrm{au}$ (where r is the spherical radius) since there is no coherent transsonic flow interior to 6 au. Top left: the gas (blue) and dust (orange) temperatures. For reference the dashed line shows the escape temperature of \citet{Owen_2012} and the dotted lines are fits to the hot isothermal section, molecular section and outer section. Top right: the gas density, with fits to the same sections. Bottom left: the proportion of H in H\textsuperscript{+} (red), H (green) and \molH{} (blue). Bottom right: the fraction of the thermochemical cooling given by each major category of cooling.}
    \label{fig:transect}
\end{figure*}

\subsubsection{Temperature, density and velocity}

Figure ~\ref{fig:rho_Tgas_deltaT} shows the gas density (left) and temperature (centre) of our fiducial model after 200 orbits at 10 au; the white dotted line in each panel indicates the sonic surface. The top panels in Fig.~\ref{fig:transect} then show the temperature and density extracted along this sonic surface; we extract these values for $r>6.2\,\mathrm{au}$ since inside this radius there is no coherent transsonic flow. The inner streamline that crosses the sonic surface at $6.2\,\mathrm{au}$ launched from the disc at $3.2\,\mathrm{au}$ at an elevation of $28^\circ$. These values are consistent with the idea that a spherical, isothermal, $10^4\,\mathrm{K}$ wind becomes transsonic at $r \gtrsim 0.5\,r_G \sim 4.5\,\mathrm{au}$ \citep{Parker_1958,Owen_2012}. While some mass-loss may be driven from within by the enthalpy \citep{Liffman_2003,Dullemond_2007,Alexander_2014}, it is generally subsonic and here becomes supersonic at $\sim 6\,\mathrm{au}$.

Qualitatively, the temperatures appear similar to those predicted by \textsc{prizmo} and \textsc{ProDiMo} in Fig.~\ref{fig:benchmark_P21} from post-processing the density field of our \citet{Picogna_2021} analogue: there is a warm, vertically extended, inner region reaching $10^4\,\mathrm{K}$, and declining temperatures outwards. Tracing along the sonic surface, Fig.~\ref{fig:transect} shows that the temperature is to a very good approximation isothermal in the inner 10-15 au before dropping suddenly. In the bottom left panel we see that this sudden drop occurs with the onset of the presence of trace amounts of molecular hydrogen in the wind, something we explore further later. Beyond this point - at around $17\,\mathrm{au}$ - out to around $120\,\mathrm{au}$, the temperature then becomes a smoothly decreasing function of radius, to which we find a power law fit of $T \propto r^{-0.97}$. This is much steeper than typically found in previous X-ray photoevaporation simulations which showed essentially isothermal winds \citep{Picogna_2021}, which we explore further in Sect. \ref{sec:explainxiT}. It is also much steeper than the values found by \citet{Nakatani_2018b}. However, it is very close to the $T_{\rm esc} \propto r^{-1}$ escape temperature predicted by \citet{Owen_2012} by assuming that nozzle flow effects are negligible in the wind, and thus balancing the gravitational force and divergence terms in a manner similar to a spherical wind \citep{Parker_1958}. We overplot the values from \citet{Owen_2012} and while they slightly overpredict the temperature at the sonic surface compared to the simulations by around 20 per cent, they are generally in good agreement. The difference could be explained by the assumptions of a spherically diverging flow or the lack of centrifugal force.

The right-hand panel of Fig.~\ref{fig:rho_Tgas_deltaT} allows us to highlight the different regimes in dust and gas temperature and how these correspond to the pinning approach of Eq.~\ref{eq:Tweighting}. At columns $\lesssim 10^{21}\,\mathrm{cm^{-2}}$, we are in the wind and $T_{\rm gas}>T_{\rm dust}$. For a column density $>10^{21}\,\mathrm{cm^{-2}}$, the two temperatures are coupled and $T_{\rm gas} \approx T_{\rm dust}$. However, since \textsc{prizmo} underpredicts the dust temperature below the photosphere, to avoid the loss of hydrostatic support to the gas, we smooth between the \textsc{prizmo} temperature and the DIAD temperature; the two diverge sufficiently little that the plotted gas and dust temperatures are the same down to a column density $10^{23}-10^{24}\,\mathrm{cm^{-2}}$, but start to diverge again beyond that, with $T_{\rm gas}>T_{\rm dust}$ since \textsc{prizmo} underestimates $T_{\rm dust}$.

We can also analyse the run of density in these regions. In the isothermal part of the wind, it scales as $\rho \propto r^{-1.01}$, similar to previous isothermal X-ray photoevaporation models \citep{Picogna_2021}. When the temperature drops at $\sim 17\,\mathrm{au}$, the density jumps in order that the pressure remains essentially smooth. It then declines again, though more shallowly, on average as $\rho \propto r^{-0.51}$. The peak number densities are therefore in the vicinity of $10^5\,\mathrm{cm^{-3}}$, more than an order of magnitude lower than previous X-ray photoevaporation models where they peaked at $\sim3\times10^6\,\mathrm{cm^{-3}}$. Even though similar $10^4\,\mathrm{K}$ temperatures are reached in these inner parts of the wind to previously, the additional cooling provided by collisional excitation of O by H means the wind can only reach them by having a much lower density (corresponding to $\xi \gtrsim 10^{-3}$, compared to $\xi \gtrsim 10^{-4}$ previously; see Sect. \ref{sec:explainxiT}) %This suggests that the overall pressure profile is similar to the isothermal region, in each case close

The panels in Fig.~\ref{fig:rho_Tgas_deltaT} are also overlaid with streamlines at regular intervals.
We start the streamlines at the location where the Bernoulli constant defined by
\begin{equation}
    B = \frac{u^2}{2} + \frac{\gamma}{\gamma-1} \frac{P}{\rho} - \frac{GM_*}{r}, 
    \label{eq:Bernoulli}
\end{equation}
(for gas velocity $u$, adiabatic constant $\gamma$, pressure $P$, density $\rho$, stellar mass $M_*$ and distance to the star $r$), becomes positive. This is an approximation - assuming an adiabatic case - for the location that the gas becomes unbound, but generally matches well the point that the streamlines start to flow away from the disc rather than trailing along its surface.

Notably, the streamlines subtend an angle of no more than $30^\circ$ with respect to the inclined wind base. This is much less than the $90^\circ$ often assumed \citep[e.g.][]{Clarke_2016}: when the temperature jump across the base is very strong, the gas is only accelerated perpendicular to the base (but not parallel to it) resulting in velocity vectors that are approximately normal to it. In this case however, the temperatures vary more smoothly from disc to wind so the perpendicular acceleration is less strong and the streamlines are more radial from the start. This means that the opening angle of the wind (the angle between the streamline and rotation axis) at the base is also larger, and ranges from $20-35^\circ$. Notably these values are centred around $30^\circ$, the minimum opening angle for a magnetocentrifugal wind \citep{Blandford_1982}. Therefore if such an angle is inferred from observations \citep[e.g.][]{Banzatti_2019,Arulanantham_2024}, caution should be exercised in immediately interpreting it as a signature of an MHD wind.

\subsubsection{Composition of the wind}
\label{sec:composition}
\begin{figure*}[ht]
    \centering
    \centering
    \includegraphics[width=\linewidth]{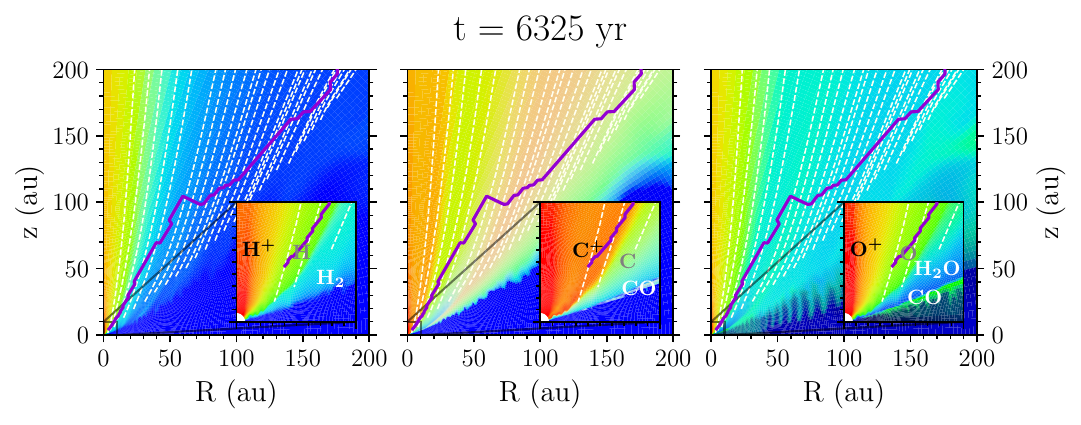}
    \caption{Three-colour images of the major carriers of hydrogen (left), carbon (centre) and oxygen (right). In each case, red colours represent ionised species (H\textsuperscript{+}/C\textsuperscript{+}/O\textsuperscript{+}), green neutral atomic species (H/C/O) and blue molecular species (H\textsubscript{2}/CO vapour+ice/\water{} vapour+ice). Therefore darker blues indicate highly molecular gas while light blues indicate gas with a significant mix of atomic and molecular species. The solid purple line in the sonic surface.}
    \label{fig:composition}
\end{figure*}

The left-hand panel of Fig.~\ref{fig:composition} shows a composite colour image representing the major H-carrying species where red represents H\textsuperscript{+}, green neutral H and blue \molH{}. As one would expect, the underlying disc is strongly molecular, while the innermost parts of the wind can become highly ionised due to EUV photons. What is perhaps more surprising is the amount of \molH{} present in the outer parts of the wind. Figure ~\ref{fig:transect} shows that its abundance rapidly grows to around 10 per cent around $\sim17\,\mathrm{au}$ and then steadily grows at the expense of neutral H until it becomes dominant outside of $50-60\,\mathrm{au}$. Indeed, there is more \molH{} than at the end of the burn-in step, despite the lower density wind conditions.
This is a surprise since \molH{} can be photodissociated by FUV \& EUV photons (while we do not have an explicit significant FUV component in our spectrum, there is still some contribution from the photosphere)\footnote{\molH{} can also be destroyed by X-ray photodissociation, but this is not included in the current version of PRIZMO.}. Moreover, our network includes the following reactions resulting in the thermal dissociation of \molH{} into atoms/ions:
\begin{align}
    \molH{}+\mathrm{H}       & \to 3\mathrm{H}  \\
    \molH{}+\mathrm{e}^{-}   & \to 2\mathrm{H}+\mathrm{e}^{-}  \\
    \molH{}+\molH{}          & \to 2\mathrm{H}+\molH{}    \\
    \molH{}+\mathrm{He}^{+}  & \to \mathrm{H}+\mathrm{H}^{+}+\mathrm{He}
    .
\end{align}

Given the wind densities suggested by our simulations, the wind will be optically thin to the photodissociating radiation.
By integrating the photodissociation cross-section across our input spectrum, we find that the photodissociation rate per \molH{} molecule is $k_{\rm PD} = 3.7\times10^{-7}\,\mathrm{s^{-1}} \frac{L_{\rm FUV}}{10^{30}\,\mathrm{erg\,s^{-1}}} \left(\frac{r}{\mathrm{au}}\right)^{-2}$, corresponding to an expected lifetime of around 25 years at $17\,\mathrm{au}$ or 2800 years at the outer edge of the wind. 
However, the self-shielding of \molH{} can help protect it once it accumulates a column of $\gtrsim 10^{14}\,\mathrm{cm^{-2}}$ \citep{Draine_1996,Richings_2014}. Given the approximate density of $10^5\,\mathrm{cm^{-3}}$ and a molecular fraction of $10^{-6}$, inside the H-\molH{} transition, the column density can only be $N \approx 10^{-6} \times 10^5\,\mathrm{cm^{-3}} \times 10\,\mathrm{au} = 10^{13}\mathrm{cm^{-2}}$. Only once the wind reaches a molecular fraction of $10^{-1}$ outside the H-\molH{} transition, can a column density of $N \approx 10^{14}\,\mathrm{cm^{-2}}$ be reached in the space of only $10^{-3}\,\mathrm{au}$. Hence, self-shielding can contribute to the steady increase in \molH{} in the wind at 10s au, but not the initial onset.

Moreover, for thermal dissociation, given the abundances in the inner disc and the $10^4\,\mathrm{K}$ temperatures involved, we expect neutral hydrogen to be the most important destruction route. This provides a thermal dissociation rate per \molH{} molecule of $k_{\rm th} = 5.2\times10^{-11}\,\mathrm{cm^{3}s^{-1}} n_{\rm H}$; given the gas densities $\sim 10^{5}\,\mathrm{cm^{-3}}$, which will be dominated by neutral H, a thermal dissociation rate of $k_{\rm th}\lesssim 6\times10^{-6}$ could be produced. This is a factor of a few higher than the photodissociation rate, so should shift the transition even further outwards, and further means that self-shielding would not protect the \molH{}.

\begin{figure*}
    \centering
    \includegraphics[width=\linewidth]{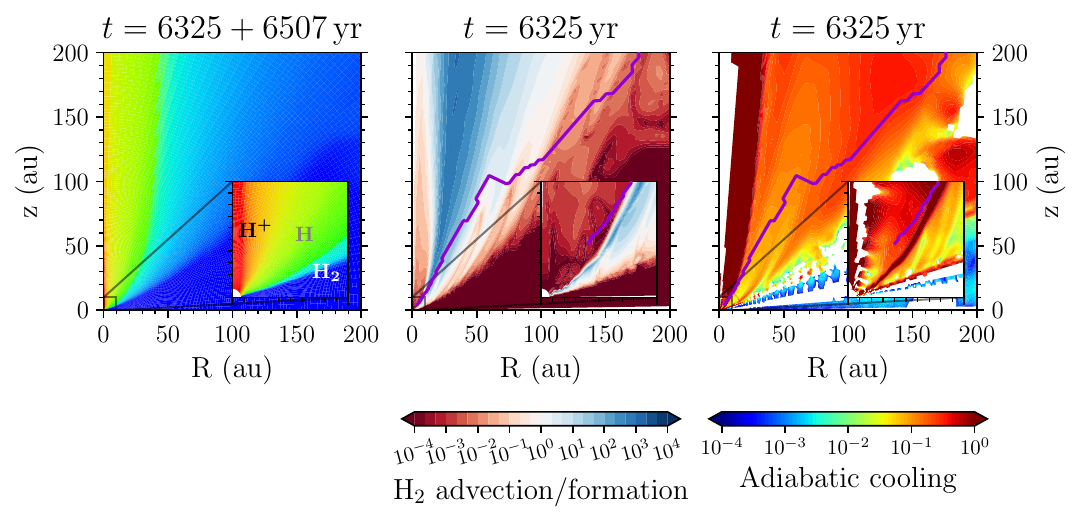}
    \caption{Effects of non-equilibrium (thermo)chemistry. Left: Composite colour image as per the right hand panel of Fig.~\ref{fig:rho_Tgas_deltaT}, after $6507\,\mathrm{yr}$ of additional evolution without hydrodynamics. Centre: Estimate of where the rate of advection of \molH{} into the wind dominates over its in situ formation on dust grains. Right: The adiabatic cooling as a fraction of the total thermochemical cooling (N.B. white regions are where PdV work instead heats the gas). In each panel, the solid purple line in the sonic surface.}
    \label{fig:advection}
\end{figure*}

Since self-shielding cannot explain the onset of \molH{}, non-equilibrium chemistry must therefore play a role. We post-process the simulation with \textsc{prizmo} for an additional $\sim 6500$ years with hydrodynamics switched off (Fig.~\ref{fig:advection}, left); we then expect that multiple photodissociation timescales will have elapsed everywhere and the \molH{} fraction will have more or less reached equilibrium.
Comparing to the left-hand panel of Fig.~\ref{fig:composition}, we can see that the colours in the wind shift away from dark and light blues towards light blues and greens - and the border between atomic and molecular winds moves outwards - indeed demonstrating an increase in neutral H as \molH{} is photodissociated and confirming that the molecular abundances are out of equilibrium.

We now estimate the hydrodynamic timescale to show that advection of \molH{} into the wind is responsible for its survival. Material at each point on the sonic surface has survived to a distance $r$ and is moving at speed $c_{\rm S}(r)$, where, as shown above, this can be approximated as the sound-speed at the escape temperature $c_{\rm S}(r) = \sqrt{\frac{GM_*}{2r}}$ \citep{Owen_2012}. This corresponds to a rate $k_{\rm ad} \sim c_{\rm S}/r \approx 1.4\times10^{-7} \left(\frac{r}{\mathrm{au}}\right)^{-1.5}$. Realistically, this is an upper limit, since the gas is initially moving subsonically.
In the right-hand panel of Fig.~\ref{fig:advection}, we also compare the rate at which \molH{} is advected into the wind in our models to the rate at which it forms on grain surfaces from purely atomic gas (the upper limit on its formation rate). This further supports our above estimate that advection is the dominant source of \molH{} in the wind on scales of 10s au. 
Note that we also have a rather large dust-to-gas ratio in the wind and lowering this may further reduce the in-situ production of \molH{} relative to its advection.
Overall we conclude that non-equilibrium chemistry is important for the composition of the wind.

\begin{figure}
    \centering
    \includegraphics[width=\linewidth]{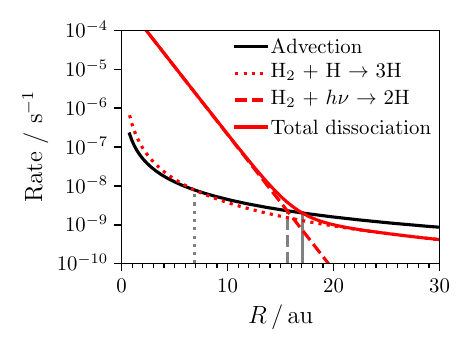}
    \caption{The rates (per \molH{} molecule) at which \molH{} is supplied to the wind by advection (black), \molH{} is photodissociated (red, dotted) and \molH{} is thermally dissociated by H (red, dot-dashed). The solid red line represents the total dissociation rate given by the sum of the other two, while the vertical grey lines mark the position where \molH{} survives against each dissociation rate.}
    \label{fig:survival}
\end{figure}

Having concluded that advection is responsible for the presence of \molH{} in the wind, in Fig.~\ref{fig:survival}, we now more carefully compare our estimate for the advective resupply timescale to the above estimate for the photodissociation rate and the expected thermal dissociation rate assuming $T(r)=T_{\rm esc}(r)$\footnote{While inside the onset of the molecular survival, the gas is actually fairly isothermal at $T\sim10^{4}\,\mathrm{K}$, if the \molH{} did survive it would be expected to stabilise $T(r) \sim T_{\rm esc}(r)$ (see later discussion), hence this is a reasonable scenario for explaining why, under those conditions \molH{} wouldn't survive if it were present (a situation which is unstable since if destroying \molH{} raises the temperature, it further increases the destruction rate).} and $n_{\rm H}(r)\approx10^{5.5}\left(r/\mathrm{au}\right)^{-1}$, in order to understand where the location of the H-\molH{} transition.
Equating the advection and photodissociation rates, we can demonstrate analytically that \molH{} would survive against photodissociation alone at
\begin{align}
    \frac{r_{\rm H2}}{\mathrm{au}}
    &= \left(\frac{3.7\times10^{-7}}{1.4\times10^{-7}}\right)^2 \left(\frac{L_{\rm FUV}}{10^{30}\,\mathrm{erg\,s^{-1}}}\right)^2 \notag \\
    &= 7 \left(\frac{L_{\rm FUV}}{10^{30}\,\mathrm{erg\,s^{-1}}}\right)^2,
    \label{eq:H2photoonset}
\end{align}
as seen in Fig.~\ref{fig:survival}.
However, at this location, thermal dissociation is dominant and so the transition does not occur here but it pushed out somewhat further. The estimated thermal dissociation rate is exponentially sensitive to temperature and therefore a much stronger function of radius than the other two rates. It falls below the photodissociation rate at $\sim16\,\mathrm{au}$; shortly thereafter at $\sim17\,\mathrm{au}$, the total dissociation rate falls below the advection rate, which is in excellent agreement with where we see the onset of \molH{} in the simulation (Fig.~\ref{fig:transect}) While photodissociation here dominates the destruction of \molH{} at its onset, albeit mildly, we note that this model has a fairly low FUV flux, and that raising this could change substantially shift the balance of destruction further towards photodissociation and push the onset of \molH{} even further out (Eq.~\ref{eq:H2photoonset}).
%This is somewhat inside of the transition seen in the simulation, but again we reiterate that this value is now a lower limit as the initially subsonic wind means that there is more time for the \molH{} to photodissociate before reaching the sonic surface.
%Although subdominant to photodissociation, thermal dissociation may also help to shift the transition outwards.

Finally, in Fig.~\ref{fig:composition}, we also show similar colour maps for C and O as for H. Here the blue regions are overall more confined to the disc, showing that molecules are less important carriers of these atoms. For example, CO is the most important molecular carrier of C, but since it is photodissociated by FUV (at a rate around 13 times faster per molecule than \molH{}) - and less abundant than \molH{} and so less able to self-shield at large radii - we only see trace quantities in the wind.
This is consistent with the results of \citet{Panoglou_2012} whose model of MHD winds suggested CO could be abundant in winds during the Class 0 phase, but highly depleted relative to \molH{} by the Class II phase.
For O, \water{} ice becomes a surprisingly larger contributor at large radii where the dust temperatures fall below the $150\,\mathrm{K}$ freeze-out temperature. In our models this happens because any water that forms freezes out and acts as a sink of O although in reality photodesorption will likely liberate the molecules which will then rapidly photodissociate (if they do not already do so in the solid state); we aim to include such processes in future work to ensure we can accurately study the outer disc. Otherwise, O is mainly neutral and atomic, and shows the same ionisation pattern as H as expected due to their near identical ionisation potentials. In contrast, C is overall somewhat more ionised since its ionisation energy is lower and thus it can be effectively ionised by highly penetrating FUV photons  

\subsubsection{Main heating \& cooling processes}
The final panel of Fig.~\ref{fig:transect} shows the fractional contribution of each process to the total thermochemical cooling.
In the inner parts of the wind, there are few molecules, and the atomic lines dominate the cooling. Once significant $\gtrsim 10$ per cent, \molH{} enters the wind, at 20-30 au, it dominates the cooling with no more than 1 per cent coming from other processes. However, we see that despite \molH{} becoming the increasingly dominant carrier of H as we move outwards, the atomic cooling slowly rises, taking over again beyond $\sim 120$ au. This occurs because the \molH{} cooling is strongly temperature dependent and beyond a certain radius where the temperature falls below $\sim 600\,\mathrm{K}$, the mid-IR lines are no longer excited. On the other hand, several atomic emission lines are present at far-IR wavelengths and so are still excited at lower temperatures. However, the lower temperatures result in lower overall cooling. 
There is also a narrow region around the onset of the molecular wind where ``chemical'' cooling - consisting of the energy lost to recombination lines and collisional dissociation of \molH{} - dominates the cooling. This is because this is the region where the wind crosses from being cooler and significantly molecular, to being hotter and fully atomic - streamlines of material do cross this border and thus \molH{} suddenly experiences hotter gas where it it rapidly dissociated. 

The ratio of the adiabatic cooling to this total is also plotted in Fig.~\ref{fig:transect}. Apart from in the narrow transition region around the onset of the molecular wind, adiabatic cooling generally contributes no more than $10$ per cent.
This is less than found by, for example, \citet{Wang_2017} because lower temperatures and higher densities typically promote two-body, line cooling processes (which are quadratic in density) at the expense of adiabatic cooling (which is only linear in density) \citep{Sellek_2022}. Moreover, \citet{Wang_2017} excluded optical emission lines and so the only active thermochemical cooling in the wind was recombination, which we do find to be everywhere less significant than adiabatic cooling in line with their results. 

The strong contribution of adiabatic cooling in the transition region occurs because gas crosses from a cooler, more molecular region where significant \molH{} cooling is possible, to where the wind is completely atomic, \molH{} becomes too rare to contribute to the cooling, and the temperatures rise. 
The gas dilutes strongly as the temperature rises - to conserve mass it must therefore also accelerate strongly, which leads to a localised increase in adiabatic cooling.

At very large radii, where line cooling becomes less effective at temperatures $T<\frac{hc}{k\lambda}$, adiabatic cooling becomes relatively more important, rising to around $30$ per cent of the thermochemical cooling rate. Again, this must imply stronger acceleration, which brings the sonic surface increasingly close to the base. This can be seen in the panels of Fig.~\ref{fig:rho_Tgas_deltaT} and means that going to large radius, the sonic surface is not a line of constant latitude but probes denser, cooler, gas, producing the increase in $\rho$ and sharper decrease in T seen beyond 120 au in Fig.~\ref{fig:rho_Tgas_deltaT}.

The role of adiabatic cooling is also shown across the whole of the wind in the right-hand panel of Fig.~\ref{fig:advection}. The same features can be seen: the steady increase towards larger radii accompanying the drop in the sonic surface towards the base amd the narrow strip of significance across the molecular-atomic transition running diagonally across the the inset. The latter appears to wider out at higher altitudes to cover most of the more highly-ionised parts of the flow (see for example the orange region in the carbon carriers in Fig.~\ref{fig:composition}) suggesting that this region is characterised by a relatively strong acceleration and poor radiative cooling.

\subsubsection{Explaining differences to previous models}
\label{sec:explainxiT}
\begin{figure*}
    \centering
    \centering
    \begin{subfigure}{0.45\linewidth}
        \includegraphics[width=\linewidth]{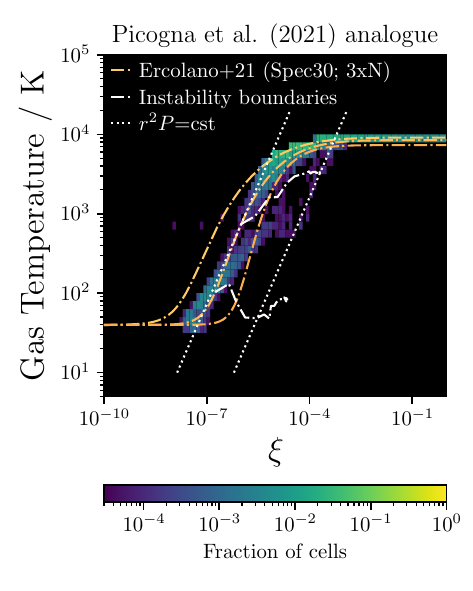}
    \end{subfigure}
    \begin{subfigure}{0.45\linewidth}
        \includegraphics[width=\linewidth]{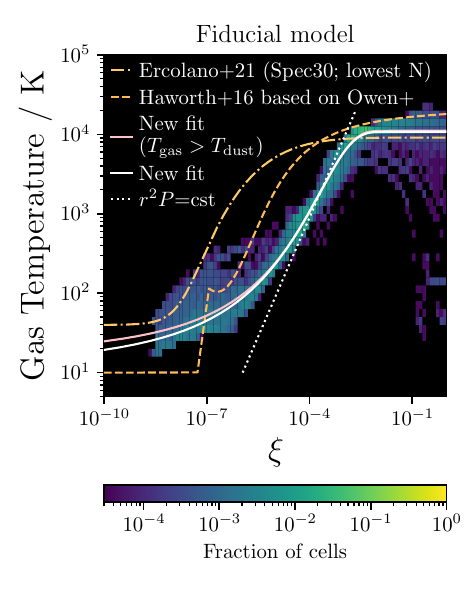}
    \end{subfigure}
    \caption{The distribution of cells in the plane of gas temperature $T$ and ionisation parameter $\xi$ for the \citet{Picogna_2021} analogue model (left) and our fiducial model (right). We only show cells which are not fixed to the initial $T_{\rm DIAD}$. Overlaid are various parametrizations of the relationships between the variables: on the left, the yellow dash-dotted lines are the parametrizations for three different column densities from \citet{Ercolano_2021}; on the right, the yellow dash-dotted line is the lowest column density \citet{Ercolano_2021} parametrization, the orange dashed line is from \citet{Haworth_2016} and the white/pink solid lines are from this work. The other white lines delineate the unstable region in the \citet{Ercolano_2021} prescription (dash-dotted) and lines of constant pressure at a given radius (dotted).}
    \label{fig:xiT}
\end{figure*}

Previous X-ray photoevaporation models calculated the gas temperatures using pre-calculated functions of the ionisation parameter $\xi=\frac{L_X}{nr^2}$. In order to understand the salient differences in our models from such works - the lower densities and the stronger temperature gradient - in Fig.~\ref{fig:xiT} we plot a 2D histogram of the temperature and ionisation parameter for our \citet{Picogna_2021} analogue \citep[which uses the $\xi-T$ relationship of ][]{Ercolano_2021} and our fiducial model. We include only cells where $T_{\rm gas} \not\equiv T_{\rm DIAD}$: for the fiducial model, this is everything with $N<10^{25}\,\mathrm{cm^{-2}}$ (Eq. \ref{eq:Tweighting}) and for the analogue model those with $N<2\times10^{22}\,\mathrm{cm^{-2}}$ and $\xi>10^{-8}$.
In both cases, we observe a tight correlation between $\xi$ and $T$ for $\xi \gtrsim 10^{-6}$; for the fiducial model somewhat more scatter occurs above the main trend for $\xi \lesssim 10^{-6}$ due to cells with $10^{21}<N\,/\,\mathrm{cm^{-2}}<10^{25}$ which are interpolated between $T_{\rm PRIZMO}$ and $T_{\rm DIAD}>T_{\rm PRIZMO}$.

For the \citet{Picogna_2021} analogue model, there are very few points in the regime $500-2000\,\mathrm{K}$, which separates the cold disc from the hot wind. Since the $\xi-T$ relationship is quite flat in the hot wind, it occupies a narrow range of temperatures and cannot achieve a significant temperature gradient.

We can understand this behaviour by considering that in order to accelerate a wind (and also in the underlying hydrostatic disc), we expect pressure to be decreasing upwards,
\begin{equation}
    \left(\frac{\partial\ln P}{\partial \theta}\right)_{r} > 0
    .
\end{equation}
Moreover, given that the wind material accelerates upwards, we also  expect an upwardly decreasing density
\begin{equation}
    \left(\frac{\partial\ln \rho}{\partial \theta}\right)_{r} > 0
    .
\end{equation}
Combining these conditions with the ideal gas law leads to the following requirement 
\begin{equation}
    \left(\frac{\partial\ln T}{\partial \ln n}\right)_{r} > -1   
    ,
\end{equation}
which, given the definition of $\xi$, we can also express as
\begin{equation}
    \left(\frac{\partial\ln T}{\partial \ln \xi}\right)_{r} < 1
    \label{eq:xiTslope}
    .
\end{equation}
Thus, we see that if the temperature increases too steeply with increasing $\xi$ then it leads to an increasing pressure, which would oppose the flow. For the $\xi-T$ relationship at each column from \citet{Ercolano_2021}, we calculate the upper and lower points where this condition is met, which we indicate with the white dash-dotted lines on Fig.~\ref{fig:xiT} - we can see the upper set is in good agreement with the deficit of points separating the locations of the cold disc and hot wind (while there are no points below the lower set since the cold disc is generally found at lower $\xi$).

For our fiducial model, the $\xi-T$ relationship in Fig.~\ref{fig:xiT} is shifted to the right with respect to those calculated by \cite{Ercolano_2021} - this implies that for the same temperature to be reached, higher $\xi$ - or equivalently lower densities - are needed. This happens, in part, due to the more efficient atomic cooling resulting from the excitation of O by H, and is seen in models even without molecular cooling. As discussed, this is important for explaining the lower density achieved in the inner $10^4\,\mathrm{K}$ part of the wind (where \molH{} does not survive and cannot contribute to the cooling).

Moreover, the relationship is much shallower than before.
The shape of the $\xi-T$ relationship is determined by the shape of the cooling function. For calculations with \textsc{mocassin} there is a lot of cooling from FIR atomic lines at low temperatures, and from optical atomic lines at high temperatures. This means that there is an intermediate temperature range where the excitation of the FIR lines has saturated, while that of the optical lines has not yet become efficient. In this regime, even a small amount of additional heating cannot be offset by additional cooling, without a large rise in the temperature in order to activate the optical lines, leading to a steep relationship between $\xi$ and T. Once the optical lines are efficiently excited, a small perturbation in temperature results in a large change in cooling and many different heating rates can be balanced in a narrow range of temperatures, leading to the roughly isothermal upper branch.

The introduction of molecular cooling, along with the presence of \molH{} in the wind due to its advection, results in a large number of lines at mid-IR wavelengths. These are able to provide cooling in that previously unstable branch, while not being so strong as to prevent additional rises in temperature entirely, thus leading to a much shallower relationship between $\xi$ and T, with maximum gradient $\frac{\partial \ln(T)}{\partial \ln(\xi)}=0.79$.
This means that inequality \ref{eq:xiTslope} is now satisfied at these temperatures, and allows them to be populated by wind material. Therefore, the full $1/r$ range of escape temperatures predicted by \citet{Owen_2012} can now be achieved. In turn, at the radii $20-120\,\mathrm{au}$ where this occurs, given that the temperature profile is fixed, the radial run of density then depends only on the slope of the $\xi-T$ profile.

We note that there is a similarity to the concept of thermal sweeping \citep{Owen_2013}, which \citet{Haworth_2016} showed happens if there is a maximum pressure along the X-ray heated $\xi-T$ curve that exceeds the pressure in the disc midplane. In this case, such a scenario is unlikely as the shallow $\xi-T$ profiles suggest that the X-ray heated gas can never exceed the pressure of the disc gas. At the very least, a very large ($\sim100 \times$) depletion in the disc gas would be necessary by the time the wind opens a cavity in order to move the disc points from $\xi\sim10^{-7}$ to $\xi\sim10^{-5}$ for thermal sweeping by hot X-ray--heated gas to then be a possible solution to the relic disc problem.

Moreover, while the microphysics of the heating and cooling are substantially different, a similar effect is known for thermal disc winds from X-rays binaries, where thermal equilibrium stability curves may be employed to find the maximum and minimum temperatures of the cool and hot branches respectively, with the stable wind conditions being those on the hot branch \citep[e.g.][]{Higginbottom_2015,Higginbottom_2017}.

The retention of a tight correlation between $\xi$ and $T$ is somewhat to be expected, since like, the case the atomic cooling previously assumed in the $\xi-T$ relationships, the \molH{} cooling happens via two-body processes. Balancing a two-body cooling rate $\Gamma = n_{\molH{}} n_{\rm coll} f(T)$ with the optically-thin X-ray photoionisation heating rate $\Lambda = \frac{n \bar{\sigma} L_{\rm X}}{4\pi r^2}$ for some spectrally-averaged cross-section $\bar{\sigma}$ allows one to find the temperature as $T=f^{-1}(\frac{\bar{\sigma}}{4\pi} \frac{L_{\rm X}}{nr^2})=f^{-1}(\frac{\bar{\sigma}}{4\pi} \xi)$ assuming that $n_{\molH{}}, n_{\rm coll} \sim n$. For a fixed spectrum and gas composition, the temperature then depends only on $\xi$.

However, since the gas composition realistically varies throughout the wind, achieving a $\xi-T$ relationship also relies on the composition (i.e., in this case, $n_{\molH{}}/n$) varying with $\xi$. This is generally true for the photoionisation level \citep[e.g.][]{Glassgold_2007}, but is less obvious for the \molH{} abundance as a result of the balance of photodissociation and advection for two reasons. Firstly, the photodissociation is driven mainly by UV, not X-ray, photons i.e. the overall dependence is on both $L_{\rm FUV}$ and $L_{\rm X}$; in this case, since the radiation field is fixed the two are directly correlated, but some caution should be exercised extrapolating to a different spectrum. Secondly, both photodissociation and advection are one-body processes, depending only on radiation field and $r$ as per Sect. \ref{sec:composition} which would suggest a relationship independent of $n$. However, at larger columns the self-shielding of \molH{} introduces a dependence of $N^{-k}$ \citep[where $k=3/4$ is often assumed, e.g.][]{Draine_1996} and thus, given the shallow density gradients seen here, a dependence on the local value of $n$. Balancing this with the advection rate described earlier, a dependence of $\frac{L_{\rm FUV}}{(nr^{5/3})^{3/4}} \sim \frac{L_{\rm FUV}}{L_{\rm X}} \xi^{3/4} L_{\rm X}^{1/4}r^{1/4} $ may result. While not solely dependent on $\xi$, given the fixed $L_{\rm X}$ and $L_{\rm FUV}/L_{\rm X}$ and weak dependence on $r$, we expect the composition to track $\xi$ closely enough to support the tight correlation seen.

Given that a tight correlation between $\xi$ and $T$ is seen despite these caveats, we can produce a $\xi-T$ fit to our simulations. We explore two fits: the first on all cells with a column density $N\leq 10^{25}\,\mathrm{cm^{-2}}$, downweighting those with $N \geq 10^{21}\,\mathrm{cm^{-2}}$ according to the weighting scheme used to interpolate between the \textsc{prizmo} and DIAD temperatures (Eq.~\ref{eq:Tweighting}); the second where we only fit the subset of those cells with $T_{\rm gas}>T_{\rm dust}$, so that we only include cells whose thermochemistry is dominated by the processes discussed above that lead to a $\xi-T$ relationship, rather than dust heating. On the other hand, for practical usage, the first fit may more smoothly transition to the dust temperatures.
We use the following fitting function \citep{Ercolano_2021}:
\begin{equation}
    \log T = d + \frac{a-d}{[1+(\log \xi/c)^b]^m}
    \label{eq:xiT}
    .
\end{equation}
The best fitting values of the parameters are given in Table \ref{tab:xiT} and show the fits on Fig.~\ref{fig:xiT}. Practically, little difference is found between the two fits except at the coolest end.
We briefly discuss the usefulness of such a fit in Sect. \ref{sec:necessity} but reserve testing whether it enables a return to a less-expensive method of determining wind temperatures for future work.

\begin{table}
    \caption{Parameters derived for the fits to the $\xi-T$ relationship}
    \label{tab:xiT}
    \centering
    \begin{tabular}{rll}
        \hline\hline
        Fit Parameter & Value (All cells) & Value ($T_{\rm gas}>T_{\rm dust}$) \\
        \hline
         $a$ & 1.04644733   & 1.19600878\\
         $b$ & -2.80200218  & -2.89745745\\
         $c$ & -0.145929345 & -0.165763115\\
         $d$ & 4.03276725   & 4.04502463\\
         $m$ & 11654.6317   & 10373.4539\\
         \hline
    \end{tabular}
    \tablefoot{See Eq.~\ref{eq:xiT} for functional form used in fits}
\end{table}

\subsection{Mass-loss rates}
\label{sec:rates}
\begin{figure*}
    \centering
    \begin{subfigure}{0.45\linewidth}
        \centering
        \includegraphics[width=\linewidth]{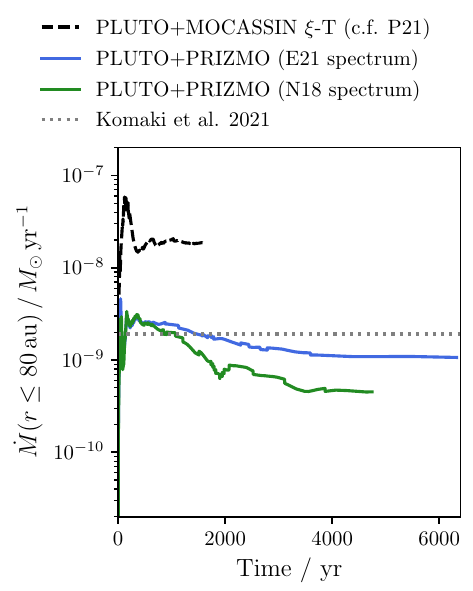}
        \caption{}
        \label{fig:Mdot_comparison_Xray}
    \end{subfigure}
    \begin{subfigure}{0.45\linewidth}
        \centering
        \includegraphics[width=\linewidth]{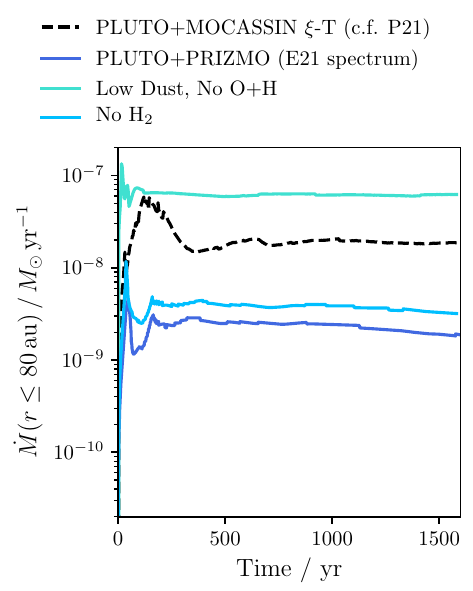}
        \caption{}
        \label{fig:Mdot_comparison_cooling}
    \end{subfigure}
    \caption{Panel a: The evolution of the mass-loss rate inside 80 au calculated using \textsc{pluto+prizmo} with the X-ray spectrum of \citet[][E21]{Ercolano_2021} (blue; 200 orbits at 10 au) and \citet[][N18]{Nakatani_2018b} (green; 150 orbits at 10 au). For comparison, we also include 50 orbits of an analogue of the \citet[][P21]{Picogna_2021} model (which uses a $\xi-T$ relationship derived from \textsc{mocassin}) as the black dashed line, and the value at $1\,{\rm M}_{\sun}$ from the fit to $\dot{M}(M_*)$ given in \citet{Komaki_2021} as the grey dotted line. Panel b: A zoom in on the evolution of the mass-loss rate inside 80 au during the first 50 orbits (at 10 au) of our fiducial simulation with the X-ray spectrum of \citet[][E21]{Ercolano_2021} (dark blue) compared with models where we turn off the cooling due to collisions between O and H and include negligible dust (turquoise) or turn off the cooling due to \molH{} (light blue). For comparison, we again include 50 orbits of the \citet[][P21]{Picogna_2021} analogue simulation as the black dashed line.}
\end{figure*}

The single key quantity from the simulations controlling the importance of photoevaporation and the lifetime of the disc is the total mass-loss rate.
In each snapshot, we calculate the cumulative mass-loss rates as a function of radius by summing the mass flux through the spherical surfaces at different radii, counting only cells where the Bernoulli constant \citep{Wang_2017} is positive and hence the material is unbound. Note that if the chosen radius is too close to the simulation outer boundary (400 au) then there may be oscillations caused by sound waves reflected off of the boundary \citep{Wang_2017,Picogna_2019}. To provide a controlled measure, and for ease of comparison to the wider literature \citep{Komaki_2021}, in Fig.~\ref{fig:Mdot_comparison_Xray} we focus on values integrated out to 80 au.

The mass-loss rate of our fiducial model settles down after just under 4000 years (or around 120 orbits at 10 au) of evolution to a value of $M(<80\,\mathrm{au})=1.06\times10^{-9}\,{\rm M}_{\sun}\,\mathrm{yr^{-1}}$. This is just over an order of magnitude lower than the \citet{Picogna_2021} analogue simulation which had settled to a mass-loss rate of $M(<80\,\mathrm{au})=1.86\times10^{-8}\,{\rm M}_{\sun}\,\mathrm{yr^{-1}}$ after 50 orbits. This suggests that the previous typical X-ray mass-loss rates of $\gtrsim 10^{-8}\,{\rm M}_{\sun}\,\mathrm{yr^{-1}}$ \citep{Owen_2010,Owen_2011,Owen_2012,Picogna_2019,Ercolano_2021,Picogna_2021} have been overestimates, consistent with evidence from population synthesis.
However, despite several differences in methodology - particularly regarding the illuminating spectrum - the mass-loss rates are now within a factor 2 of those calculated for a solar-mass star by \citet{Komaki_2021}, who also performed on-the-fly thermochemistry with hydrodynamics. This is an encouraging step towards a convergence of mass-loss rate estimates between works and in stark contrast to disagreement by several orders of magnitude seen previously.

We note however, that in our model there is still significant mass-loss beyond $80\,\mathrm{au}$: fitting out to the observed drop-off in the surface density profile at $\sim 160\,\mathrm{au}$ suggests a total value 3-4 times higher at $\dot{M}_{\rm tot} = 4.3178\times10^{-9}\,{\rm M}_{\sun}\,\mathrm{yr^{-1}}$. This is still $\sim9$ times lower than the total rate found by \citet{Picogna_2021} for the same stellar mass and spectrum.

\subsubsection{Dependence on spectrum}
\label{sec:spectrum}
Due to the strong dependence of the cross-section for photoionisation on photon energy, X-ray photoionisation heating is very sensitive to the shape of the X-ray spectrum \citep{Sellek_2022}. The results shown here - for example, the order of magnitude of the mass-loss rate and the presence of a molecular photoevaporative wind - are qualitatively similar to those presented in \citet{Nakatani_2018b,Komaki_2021}. However, these works argued that the contribution of X-rays to driving the wind was negligible, unlike the results of our simulations.
Therefore we consider if the different shapes of the X-ray spectra employed could be driving the difference.

The TW Hya spectrum \citep{Nomura_2007} used by \citet{Nakatani_2018b,Komaki_2021} (see Fig.~\ref{fig:spectra}) clearly lacks significant amounts of flux in the soft X-rays compared to that used by \citet{Ercolano_2021}, and the EUV is almost entirely absent. This is despite us normalising the two spectra to the same total X-ray luminosity ($2\times10^{30}\,\mathrm{erg\,s^{-1}}$ between $0.5-5\,\mathrm{keV}$) and the fact that TW Hya has a notably ``soft'' X-ray spectrum \citep{Kastner_2002} compared to other T Tauri stars \citep[which here manifests as the faster drop-off at $E>1\,\mathrm{keV}$ due to the lack of the highest-temperature plasma component;][]{Nomura_2007}.
The difference at the soft end arises because in X-ray spectra as observed at the Earth, these energies are attenuated due to absorption (primarily by neutral hydrogen) in the foreground interstellar medium along the line of sight\footnote{Although in principle, there could be some contribution from the circumstellar material - including any wind - TW Hya is close to pole-on so this contribution should be negligible.}. The \citet{Nomura_2007} spectrum is a model for this observed flux - and thus includes attenuation by their estimated column of $N_{\rm H}=2.7\times10^{20}\,\mathrm{cm^{-2}}$ - while the \citet{Ercolano_2021} spectrum is a model for the intrinsic emission of the star - and is corrected for this attenuation. The latter is what the wind receives and is thus the more physical model.

\begin{figure}
    \centering
    \includegraphics[width=\linewidth]{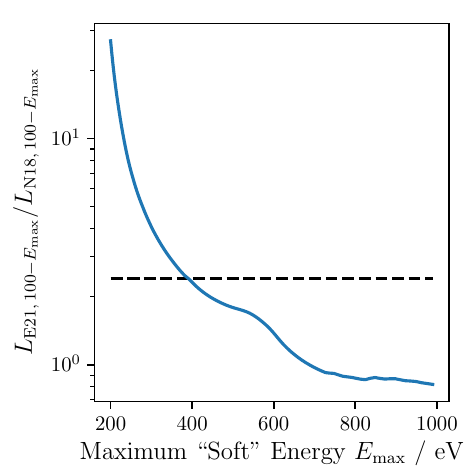}
    \caption{The ratio of the soft X-ray luminosity of the two X-ray spectra investigated in this work, as a function of the maximum energy deemed to be soft. The black dashed line shows the ratio of the two mass-loss rates for comparison.}
    \label{fig:Lsoft_ratio}
\end{figure}

Figure \ref{fig:Mdot_comparison_Xray} shows that the impact of using the observed spectrum instead of an intrinsic spectrum is to lower the mass-loss rate by a factor 2-3. This indeed suggests that the mass loss is driven by soft X-rays as usually argued \citep{Ercolano_2009}. Calculating the ratio of the luminosites of the two spectra integrated from $100\,\mathrm{eV}$ to different upper limits $E_{\rm up}$ (Fig.~\ref{fig:Lsoft_ratio}), we find that this ratio equals the ratio of mass-loss rates for $E_{\rm up}\approx400\,\mathrm{eV}$. Under the typical assumption that the mass-loss rates are approximately linear with the appropriate X-ray luminosity \citep{Owen_2012}, this suggests that the photoevaporation in our simulations is being driven by X-rays with energies $100-400\,\mathrm{eV}$.

\subsubsection{Dependence on microphysics}
Since our fiducial spectrum is identical to that used by \citet{Ercolano_2021,Picogna_2021}, this cannot be a reason for the difference in mass-loss rates between this work and the previous simulations.

Instead, the same microphysics that results in the drastically different temperature profile can also explain the lower mass-loss rates. Figure \ref{fig:Mdot_comparison_cooling} shows the mass-loss rate for models where we run the hydrodynamics for a short time with either the dust-to-gas ratio reduced to a negligible value and the collisions between neutral H and O disabled - since this was a particularly effective combination for changing the temperature in our benchmarking - or the cooling from \molH{} excluded - since this dominates over most of the wind in our fiducial model.

In the latter case we see that neglecting \molH{} cooling boosts the mass-loss rate by only a factor $\sim2$. While important, this cannot fully explain the difference to previous models.
Instead, in the former case, we find a much more dramatic difference, with the mass-loss rate being more than an order of magnitude higher. As shown in Fig.~\ref{fig:benchmark_P21}, this combination of reduced dust and no excitation of O by neutral H significantly reduces the cooling, which there resulted in a hot, approximately isothermal wind to persist. Compared to our fiducial model then, gas of a given density can reach a high temperature, meaning that the escape temperature can be reached for higher density gas, leading to a higher density wind.
The fact that these high mass-loss rates are even higher than those seen in previous X-ray simulations likely results from weaker atomic cooling due to lower collisional excitation rates by electrons \citep{Bell_1998}, as well as the smaller collections of lines we included (for example by neglecting S due to its likely strong depletion).

Therefore, the lack of dust and neutral colliders in \textsc{mocassin} are the most likely reason for the high-mass loss rates seen in previous X-ray photoevaporation simulations. Since our dust-to-gas ratio is likely too high (see Sect. \ref{sec:dustentrainment}), then the latter of these is the more important reason. However, the advective survival of \molH{} then allows it to become the most important coolant limiting the mass-loss rates, something that could not be captured without the on-the-fly thermochemistry.

\subsubsection{Mass-loss profile}
\label{sec:profile}
\begin{figure*}
    \centering
    \begin{subfigure}{0.45\linewidth}
        \includegraphics[width=\linewidth]{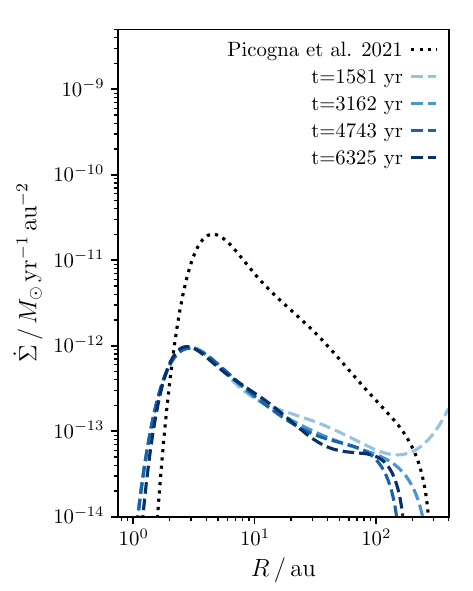}
    \end{subfigure}
    \begin{subfigure}{0.45\linewidth}
        \includegraphics[width=\linewidth]{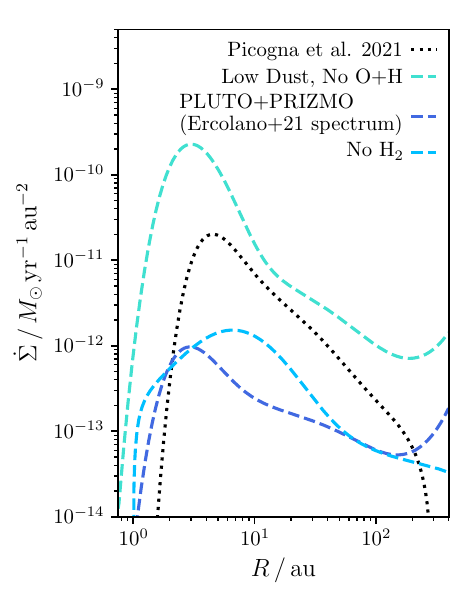}
    \end{subfigure}
    \caption{The mass-loss rate per unit area as a function of distance to the star. Left: different times (from light to dark blue with increasing time) for our fiducial model compared to those from \citet{Picogna_2021} for the same stellar mass and X-ray spectrum (black). Right: comparing the fiducial model (dark blue) with those where we switch off certain areas of microphysics in order to test their impact.}
    \label{fig:Mdot_profile_evolution}
\end{figure*}

In Fig.~\ref{fig:Mdot_profile_evolution} we show fits to the mass-loss profile in our fiducial simulations taken every $50$ orbits at 10 au. We determine these in the same way as before: by calculating the cumulative mass-loss profile, $\dot{M}(<r)$, fitting a sixth-order polynomial in log-log space and differentiating:
\begin{equation}
\begin{split}
    \log\left(\frac{\dot{M}(<R)}{\dot{M}_{\rm tot}}\right) = a \log(R)^6 + b \log(R)^5 + c \log(R)^4 + d \log(R)^3 \\ + e \log(R)^2 + f \log(R) + g,
    \label{eq:Mdotcumfit}
\end{split}
\end{equation}
\begin{equation}
    \dot{\Sigma} = \frac{1}{2 \pi R} \frac{{\rm \partial} \dot{M}(<R)}{{\rm \partial} R}
    \label{Sigmadotfit}
    .
\end{equation}
The parameters of the fit for the final snapshot are given in Table \ref{tab:profilefit}; $\dot{M}_{\rm tot}$ is measured out to the point where the fit predicts a negative mass-loss rate per unit area (which we use as the outer edge of the wind) - and is the value that should be used in the fit - while $\dot{M}_{\rm true}(<80\,{\rm au})$ is the equivalent measurement out to 80 au, and $\dot{M}_{\rm fit}(<80\,{\rm au})$ is the value at 80 au predicted by Eq.~\ref{eq:Mdotcumfit}.

\begin{table}
    \caption{Parameters derived for the fit to the mass-loss profile}
    \label{tab:profilefit}
    \centering
    \begin{tabular}{rl}
        \hline\hline
        Fit Parameter & Value \\
        \hline
         $\dot{M}_{\rm tot}$ & $4.3178\times10^{-9}\,{\rm M}_{\sun}\,\mathrm{yr^{-1}}$ \\
         $a$                 & $-1.2108$ \\
         $b$                 & $9.6815$ \\
         $c$                 & $-31.226$ \\
         $d$                 & $52.293$ \\
         $e$                 & $-48.475$ \\
         $f$                 & $25.093$ \\
         $g$                 & $-7.6443$ \\
         \hline
         $\dot{M}_{\rm fit}(<80\,{\rm au})$ & $1.50\times10^{-9}\,{\rm M}_{\sun}\,\mathrm{yr^{-1}}$ \\
         $\dot{M}_{\rm true}(<80\,{\rm au})$ & $1.06\times10^{-9}\,{\rm M}_{\sun}\,\mathrm{yr^{-1}}$ \\
        \hline
    \end{tabular}
    \tablefoot{
    The form of the profile is described by Eq.~\ref{eq:Mdotcumfit}. The bottom two values give, for reference, the cumulative mass loss rate inside 80 au predicted from the fit and that measured directly in the simulation (as used in Sect. \ref{sec:rates}).
    We recommend using the full precision given here since the parameters are fits to exponents which can result in truncation errors compounding to give a poor fit at large radii.}
\end{table}

As the simulation evolves, most of the change can be seen in the outer regions of the wind, while the innermost parts are unchanged (due to the freezing of the simulation inside $r\approx17\,\mathrm{au}$). At the earliest time shown the outer parts of the simulation have not settled, and so an unphysical increase towards the outer boundary is seen. As the outer regions settle, the outer radius decreases to $\sim 160\,\mathrm{au}$ up to around 150 orbits, before increasing slightly again to $180\,\mathrm{au}$ by the final plotted snapshot. This suggests that the outer regions of the simulation have mostly stabilised, although running for a bit longer may demonstrate additional mass-loss from the outermost parts as the disc settles.

Aside from maybe at its very innermost edge, the profile is everywhere much lower than seen in the previous models by \citet{Picogna_2021} - the additional microphysics included here compared to previously lowers the mass-loss rate from the disc everywhere. The slope of the profile is also a little shallower than previously due to the much less steep dependence of $\rho$ on r. The inward shift of the peak and inner edge is likely due to slightly higher maximum temperatures in the wind, resulting in unbound material closer to the star. These temperatures are reached since the lower column densities are lower than the lowest slabs considered by \citet{Ercolano_2021,Picogna_2021}.

When the dust-to-gas ratio is turned down to a negligible value and the collisions between neutral H and O are disabled, the mass-loss rate is increased by roughly the same factor $\gtrsim 10$ everywhere (though is slightly higher close to the star where temperatures are hotter); the reduction in mass-loss caused by this process is not isolated to any one location in the wind and is in line with our statement that the reduced cooling allows denser gas to reach the escape temperature.
On the other hand, when cooling due to \molH{} is disabled, the mass-loss rates are fairly similar around $R\lesssim 3\,\mathrm{au}$ and at $R\gtrsim 50\,\mathrm{au}$.
Note that smaller radii the mass-loss rates are still much smaller than previous works because the density of the wind has to be lower to overcome the additional heating from collisional excitation of O by neutral H and still reach the $\sim10^4\,\mathrm{K}$ escape temperature.
The greatest difference is seen around $R = 10 \mathrm{au}$ where \molH{} starts to survive into the wind and where the wind is warm enough to excite the \molH{} lines leading to additional cooling. At smaller radii there is no \molH{} in the wind to provide cooling, while outside 50 au the wind is too cold to excite it.

\section{Discussion}
\label{sec:discussion}
\subsection{Necessity of this method}
\label{sec:necessity}
There remains the question of whether we needed to adopt such as expensive alternative method of calculating the thermochemistry.
It has previously been argued that a $\xi-T$ approach neglects the adiabatic cooling due to expanding gas and thus risks double counting the input of energy \citep{Wang_2017}. However, we find that this term is nearly everywhere subdominant to the atomic and/or molecular line cooling and radiative thermal equilibrium is a good approximation\footnote{The effects of adiabatic cooling can be seen in Fig.~\ref{fig:xiT} as the small number of points with $\xi>10^{-3}$ and $T<10^4\,\mathrm{K}$ and correspond to bound cells at small radii near the rotation axis.}. This is supported by the close correlation between the ionisation parameter $\xi$ and gas temperature $T$ (as shown in Sect. \ref{sec:explainxiT}), which demonstrates that this can remain an appropriate parametrization of the underlying microphysics.

However, we note that over much of the wind, we found cooling to be dominated by \molH{}, the abundance of which is out of chemical equilibrium due to advection.
\citet{Komaki_2021}, who derive similar photoevaporation rates also see partially molecular winds, and the importance of advection for the survival of molecules into the wind has also previously been argued for photoevaporation \citep{Wang_2017,Nakatani_2018a,Nakatani_2018b} and MHD winds \citep[][]{Panoglou_2012,Wang_2019}.
This is the most important impact of the hydrodynamics on the temperature structure in our work: such an effect could not be captured by the sort of static slab models used to calibrate the $\xi-T$ relationships before, thus suggesting it is necessary to perform the chemistry on the fly. However, since it does seem that despite this, a close $\xi-T$ relationship can be found, and since it has previously been demonstrated that the $\xi-T$ approach can be adapted to varying chemical abundances \citep{Wolfer_2019} - as well as column densities \citep{Picogna_2019} and different spectra \citep{Ercolano_2021} - a future intermediate possibility could be to produce a set of $\xi-T$ relationships at differing \molH{} abundances to simplify at least the temperature calculation.

\subsection{Dust entrainment}
\label{sec:dustentrainment}
Dust entrainment in winds is an important question both as a sink of mass, and as a potentially observable tracer of the wind \citep[][]{Owen_2011b,Franz_2022a,Franz_2022b}.
Both the delivery of dust to the base of the wind \citep{Hutchison_2021,Booth_2021}, and the ultimate trajectory in the wind \citep[e.g.][]{Franz_2020} can play a role in this picture. Moreover, the level of dust may affect the thermochemistry in the wind as discussed in Sect. \ref{sec:benchmarks_Xray}; we therefore consider how self-consistent our simulations are.

\begin{figure*}
    \centering
    \includegraphics[width=\linewidth]{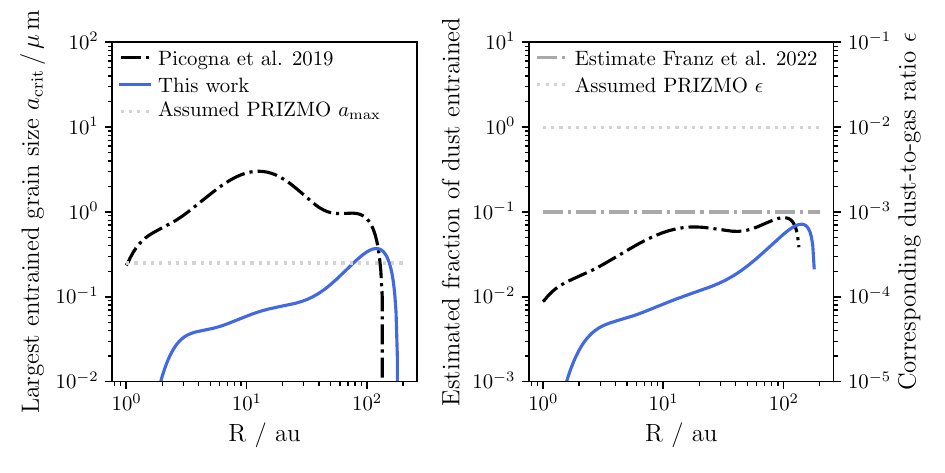}
    \caption{Left: the maximum entrained grain size in the wind according to the prescription of \citet{Booth_2021} for our fiducial model (blue), compared to that expected previous model of \citet{Picogna_2019}. Right: an estimate for the resulting fraction of dust mass/dust-to-gas ratio in the wind. The grey lines indicate the values used in this model.}
    \label{fig:dustentrainment}
\end{figure*}

Using our fitted mass-loss profile to the fiducial model in Eq.~25 of \citet{Booth_2021}, we show the maximum entrained grain size as a function of radius in Fig.~\ref{fig:dustentrainment}. The values lie in the range of $0.04-0.4\,\mu\mathrm{m}$ and are typically 1-2 orders of magnitude below those that \citet{Booth_2021} calculated using a previous X-ray photoevaporation model \citep{Picogna_2019}, in line with our mass-loss rate which is approximately an order of magnitude smaller. This range straddles our assumed maximum grain size of $0.25\,\mu\mathrm{m}$, suggesting that this is a reasonable value, though we also note that the entrained size is smaller in the inner disc compared to the outer.

By assuming that the maximum dust size in the disc is set by fragmentation according to Eq.~3 of \citet{Birnstiel_2012} with $\alpha=10^{-3}$ and $u_{\rm frag}=10^2\,\mathrm{cm\,s^{-1}}$, and assuming a \citet{Lynden-Bell_Pringle_1974} disc profile with $M_{\rm D}=0.1\,{\rm M}_\sun$, $R_{\rm C}=100\,\mathrm{au}$ and $T_{\rm 1 au}=270\,\mathrm{K}$, we can estimate the maximum grain size at each radius in the disc $a_{\rm frag}$. For a \citet{Mathis_1977} grain size distribution, then the entrained fraction is $f_{\rm ent} = \left(\frac{a_{\rm max}}{a_{\rm frag}}\right)^{1/2}$. For the \citet{Picogna_2019} mass-loss profile, this estimate suggests up to $\sim10$ per cent entrained mass (or a dust-to-gas ratio of $10^{-3}$), consistent with the modelling by \citet{Franz_2022a}. Meanwhile, our mass-loss profile suggests values up to 5 times lower, largely in the range $\epsilon = 10^{-4}-10^{-3}$. This suggests that the dust-to-gas ratio we assume is $10-100$ times too high, and we should explore lower values in future models.

\subsection{Consequences for disc evolution \& population synthesis}
\label{sec:evolution}
One of the major usages of photoevaporation prescriptions such as we derive in Sect. \ref{sec:profile} is to include in disc evolution models, often for the purposes of population synthesis. One limitation of the previous X-ray rates is that they needed fairly careful selection of the parameter space in order to get long-enough lived discs.
For example, several works find it difficult to get discs that live for $3\,\mathrm{Myr}$ or more (and therefore survive to the ages of older regions such as Upper Sco) when using the median X-ray luminosities \citep{Emsenhuber_2023,Appelgren_2023}.
Similarly, \citet{Somigliana_2020} predict a ``knee'' in the plane of accretion rate vs disc mass that is not observed, which limits photoevaporation to low values, and \citet{Sellek_2020b} argue that when dust is included, high photoevaporation rates for all stars would not allow discs to reach such low dust masses as are observed while remaining accreting.

One way to understand this is with the so-called ``UV-switch'' model \citep{Clarke_2001}: once the photoevaporation rate exceeds the accretion rate, accretion can no longer resupply the inner disc as all of the material it brings in is immediately ejected in the wind. This results in the opening of a gas cavity.
When the accretion rates are higher than $10^{-8}\,{\rm M}_{\sun}\,\mathrm{yr^{-1}}$, they are comparable to the median accretion rate \citep{Manara_2023} and consequently a disc may have very little chance to evolve before reaching this criterion. The lower values that we find thus delay the onset of the rapid disc dispersal.

This has motivated people to question which rates best fit the distribution of accretion rates. On the one hand, \citet{Alexander_2023} showed that for $0.3 \leq M_*/{\rm M}_{\sun} \leq 1.2$ if the mass-loss rate exceeds $\sim10^{-9}\,{\rm M}_{\sun}\,\mathrm{yr^{-1}}$ then the process of gap opening, inner disc dispersal and cessation of accretion happens so quickly that there would be too few discs with low accretion rates to match the observed distribution.
On the other hand, \citet{Ercolano_2023} argue that if one constructs a population of discs to fit the disc fraction with age (i.e. the lifetime distribution) under the scenarios of EUV-, X-ray- and FUV- driven photoevaporation, the best fit to the distribution of low accretion rates is with X-ray photoevaporation \citep[for which they used the higher rates of][]{Picogna_2021}. This happens because EUV photoevaporation requires more tenuous discs in order for them to be dispersed in the same time, and therefore lower accretion rates.

\begin{figure*}
    \centering
    \includegraphics[width=\linewidth]{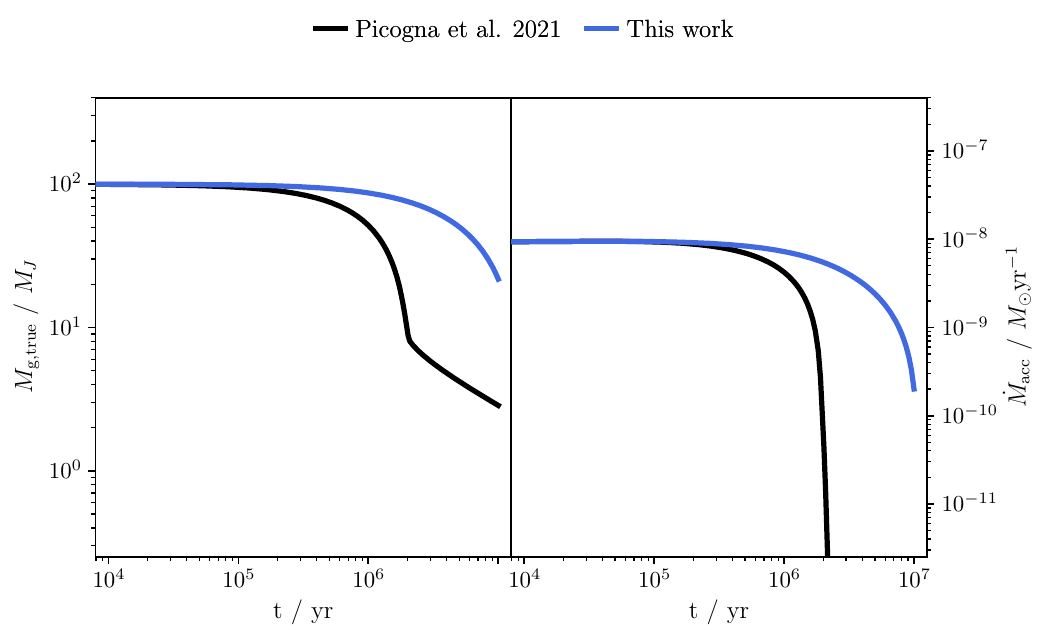}
    \caption{Disc gas mass (left) and accretion rate (right) over the course of 10 Myr for an example $100\,{\rm M_J}$, $100\,\mathrm{au}$ disc around a solar-mass star undergoing viscous evolution ($\alpha=10^{-3}$) and photoevaporation using the mass-loss prescription of \citet{Picogna_2021} (black) or from this work (blue).}
    \label{fig:discevolution}
\end{figure*}

Nevertheless, our lower mass-loss rates are a step in the right direction in terms of allowing photoevaporation to be consistent with the distributions of disc accretion rates and lifetimes found most commonly in the literature.
We illustrate this in Fig.~\ref{fig:discevolution}, where we compare the evolution of identical $100\,{\rm M_J}$, $100\,\mathrm{au}$ discs around solar-mass stars with undergoing viscous evolution ($\alpha=10^{-3}$) and photoevaporation according to the prescriptions of \citet{Picogna_2021} and Sect. \ref{sec:profile} of this work. Since less mass is removed, our prescription results in larger gas masses at all times, and allows the disc to keep accreting at detectable levels for the full 10 Myr that we simulated. Moreover, around 10 per cent of the mass is left by the time of the cavity opening, suggesting that thermal sweeping would not kick in to prevent the existence of massive but non-accreting ``relic discs''.

A final consequence of lowering the internal photoevaporation rate is that the relative importance of other mass-loss processes, such as external photoevaporation may be increased in a given environment. This may, for example, cap the possible disc lifetimes and lead to fewer cases of inside-out disc clearing \citep{Coleman_2022}.

\subsection{Observables}
\label{sec:observables}
The line emission from winds is collisionally excited and, consequently, sensitive to temperature. Depending on the excitation temperature, the reduction in the gas temperature at large radii in the winds is therefore also likely to reduce the population of the upper levels of the lines. For example, the commonly studied [\ion{Ne}{II}] $12.81\,\mu\mathrm{m}$ line has an excitation temperature of $1123\,\mathrm{K}$ and therefore would not be emitted beyond about $60\,\mathrm{au}$. However, in previous models, this line mainly originated inside $10-20\,\mathrm{au}$ anyway \citep{Ercolano_2010,Ercolano_2021}, suggesting that any effect would be small. However, given that the velocities in the wind are of order $c_{\rm S}$, the cooler wind will also be slower, and thus present high-spectral-resolution line profiles that are less blue-shifted and narrower \citep{Ballabio_2020}.

The lower densities - and shallower radial run of density may also have important consequences. \citet{Sellek_2024a} showed that for appropriate column densities, the absorption of soft X-rays in the inner wind can lead to [\ion{Ar}{II}] emission becoming more compact than the [\ion{Ne}{II}] emission. In our models, more absorption would occur at larger radii, making such a configuration harder to achieve. Moreover, having shallower density profiles will produce narrower line profiles \citep{Ballabio_2020}.

Finally, the fact that molecules - particularly \molH{} - survive further into the wind (and closer to the star) than otherwise expected due to advection suggests that photoevaporation could also be traced by molecular lines, such as the $2.12\,\mu\mathrm{m}$ line \citep{Gangi_2020}. This was already shown in \citet{Rab_2022} for photoevaporative winds using a simpler post-processing approach for the thermochemistry. However, as noted in \citet{Rab_2022}, and as shown in this work, such an approach has to be used with care. Since the material is seen further along the streamlines, it will have higher velocities than static post-processing would suggest and, therefore, produce broader and more highly blue-shifted lines.
We intend to use our model to produce synthetic observables in a future study in order to quantify more precisely the importance of advection for interpreting molecular wind tracers.

\section{Conclusions}
\label{sec:conclusions}
In this work we have presented improvements to the thermochemistry code \textsc{prizmo} \citep{Grassi_2020} - designed for use on-the-fly with hydrodynamics - and benchmarked its performance for describing the temperatures of photoevaporative winds in different regimes against previous works and other codes. We have then demonstrated the first science application to the problem of the thus-far divergent mass-loss rates for photoevaporation in the literature, focusing on the X-ray regime.
By a) designing our approach to be sufficiently flexible and comprehensive to capture thermochemistry in ionised, neutral atomic and molecular environments at a range of temperatures and b) testing spectra used in different previous works in a controlled manner, we have ensured that we can confidently isolate the root of the differences we find and probe the causes of the different claims in the literature.
In particular, we find that
\begin{enumerate}
    \item The temperatures of X-ray driven photoevaporative winds, which are very weakly ionised, can be greatly overestimated if neutral H is not included as a collider to excite atomic forbidden line emission (Fig.~\ref{fig:benchmark_P21}).
    \item The near-isothermal temperatures seen in previous X-ray photoevaporation models can be understood as a result of the very steep relationships between the temperature and ionisation parameter $\xi$ used therein leading to a dearth of cells at unstable temperatures $\sim1000\,\mathrm{K}$ (Fig.~\ref{fig:xiT}). In turn this occurs because of the bimodal distribution of atomic emission lines between the optical and far-IR regimes; molecular cooling in the mid-IR stabilises the intermediate temperatures leading to shallower relationship. As a result, the new simulations self-regulate to achieve the $1/r$ escape temperature predicted by \citet{Owen_2012} (Fig.~\ref{fig:transect}).
    \item Adiabatic cooling of the gas is only dominant in a very narrow region along the \molH{} dissociation front (Fig.~\ref{fig:transect}), and very low-density gas along the rotation axis. However, it does become relatively more important at larger radii where the gas is cooler as the excitation of line cooling becomes harder.
    \item As a result of the additional cooling mechanisms discussed above, the escape temperature can only be reached in less-dense gas; this lowers the mass-loss rates of X-ray photoevaporation for a solar mass star by over an order of magnitude to $\sim 10^{-9}\,{\rm M}_{\sun}\,\mathrm{yr^{-1}}$ (Fig.~\ref{fig:Mdot_comparison_Xray}). The mass-loss is reduced everywhere by a similar amount compared to previous works (Fig.~\ref{fig:Mdot_profile_evolution}).
    \item The winds are driven by soft X-rays; use of an observationally-derived spectrum without correcting for the foreground attenuation of these energies can lead to underestimates of the X-ray contribution to photoevaporation (Fig.~\ref{fig:Mdot_comparison_Xray}).
    \item The winds contain a significant amount of molecular hydrogen; advection of material into the wind - i.e. non-equilibrium chemistry - is critical for its existence (Fig.~\ref{fig:advection}). This will be important for matching the extent and velocity of line diagnostics probing molecular winds.
\end{enumerate}

The lower mass-loss rates for X-ray photoevaporation, which we have understood to result from the additional cooling provided, in particular, by excitation of O by neutral H, help to resolve long-standing tensions in the field of photoevaporation simulations and move us towards a consensus on the expected photoevaporation rates.

These results have significant implications for the evolution and demographics of protoplanetary discs. As a result of the lower mass-loss rates, the lifetime of discs can be considerably longer than expected previously and they should be able to reach lower accretion rates before undergoing rapid cavity opening and dispersal. This helps resolve tensions with works on disc demographics which generally imply lower photoevaporation rates than the previous X-ray models suggested.
Future work will need to explore the parameter space, to see how these rates depend on the amount of FUV irradiation as well as the stellar mass, which will enable the production of a general prescription to be used in disc and planet population synthesis models.

\section*{Data availability}
The files needed to setup and run the fiducial model, as well as its final output, are available from Zenodo: \url{https://doi.org/10.5281/zenodo.10891357}. The version of \textsc{pluto} with the \textsc{prizmo} interface is available from GitHub: \url{https://github.com/AndrewSellek/PLUTO_PRIZMO/tree/chemistry}, while \textsc{prizmo} itself is available from GitHub: \url{https://github.com/tgrassi/prizmo}.

\begin{acknowledgements}
        We thank the reviewer, James Owen, for his detailed feedback which helped us to clarify various aspects of the microphysics and technical approach.
        We are grateful to Lile Wang for sharing the density and temperature grids from \citet{Wang_2017} used in the benchmarking, Catherine Espaillat for providing the DIAD models and Riouhei Nakatani for useful discussions about the comparison between our model and his and for sharing details of his X-ray spectrum. ADS also thanks James Matthews for useful conversations about the similarities to thermal winds from X-ray binaries.

        ADS thanks the Science and Technology Facilities Council (STFC) for a Ph.D. studentship for project 2277492 as part of Training Grant ST/T505985/1 and was also supported by funding from the European Research Council (ERC) under the European Union’s Horizon 2020 research and innovation programme (grant agreement No. 1010197S1 MOLDISK).
        CJC acknowledges support from the Science \& Technology Facilities Council (STFC) Consolidated Grant ST/W000997/1 and European Union's Horizon 2020 research and innovation programme No 823823 (DUSTBUSTERS) under the Marie Sklodowska-Curie grant agreement.
        We acknowledge the support of the Deutsche Forschungsgemeinschaft (DFG, German Research Foundation) Research Unit ``Transition discs'' - 325594231. This research was supported by the Excellence Cluster ORIGINS which is funded by the Deutsche Forschungsgemeinschaft (DFG, German Research Foundation) under Germany's Excellence Strategy - EXC-2094 - 390783311. CHR is grateful for support from the Max Planck Society.

        This work was performed using resources provided by the Cambridge Service for Data Driven Discovery (CSD3) operated by the University of Cambridge Research Computing Service (www.csd3.cam.ac.uk), provided by Dell EMC and Intel using Tier-2 funding from the Engineering and Physical Sciences Research Council (capital grant EP/T022159/1), and DiRAC funding from the Science and Technology Facilities Council (www.dirac.ac.uk).

        This work also benefited from the Core2disk-III residential program of Institut Pascal at Universit\'e Paris-Saclay, with the support of the program ``Investissements d’avenir'' ANR-11-IDEX-0003-01.
\end{acknowledgements}

% WARNING
%-------------------------------------------------------------------
% Please note that we have included the references to the file aa.dem in
% order to compile it, but we ask you to:
%
% - use BibTeX with the regular commands:
%   \bibliographystyle{aa} % style aa.bst
%   \bibliography{Yourfile} % your references Yourfile.bib
%
% - join the .bib files when you upload your source files
%-------------------------------------------------------------------

\bibliographystyle{aa}
\bibliography{biblio}%,bibilio_extra}

%-------------------------------------------------------------
%               Appendices have to be placed at the end, after
%                                        \end{thebibliography}
%-------------------------------------------------------------

\begin{appendix}

\section{Fits to OI collisional de-excitation data}
\label{appendix:Lique}
We perform our own fits to the tabulation of the de-excitation rates of \ion{O}{I} by H and \molH{} from \citet{Lique_2018}, in the form
\begin{equation}
    k_{\rm ul} = (a \times T_2^{b + c \ln T_2} + d \exp(-e/T_2)) \times 10^{-10}~\mathrm{cm^3~s^{-1}}
    \label{eq:Lique}
    ,
\end{equation}
where $T_2 = \frac{T}{10^2\,\mathrm{K}}$. The values of the parameters are reported in Table \ref{tab:Lique}.

\begin{table*}[h]
    \caption{Parameters derived for the fits to collisional de-excitation rates of \ion{O}{I} by H and \molH{}}
    \label{tab:Lique}
    \centering
    \begin{tabular}{lllllllll}
    \hline\hline
    Collider & Transition & Energy Levels & Wavelength & a & b & c & d & e \\
     &  &  & $(\mu\mathrm{m})$ &  &  &  &  &  \\
    \hline
    H
    &\term{3}{P}{0}$\to$\term{3}{P}{1} & $E_2 \to E_1$ & 145
    & 1.19 & 0.160 & 0.072 & 5.03 & 18.9 \\
    &\term{3}{P}{1}$\to$\term{3}{P}{2} & $E_1 \to E_0$ & 63
    & 1.37 & 0.348 & 0.018 & 2.96 & 3.77 \\
    &\term{3}{P}{0}$\to$\term{3}{P}{2} & $E_2 \to E_0$ & 44
    & 1.03 & 0.304 & 0.033 & 3.08 & 5.18 \\
    \hline
    para-\molH{} 
    &\term{3}{P}{0}$\to$\term{3}{P}{1} & $E_2 \to E_1$ & 145
    & 0.00719 & 0.624 & 0.051 & 0.338 & 36.2 \\
    &\term{3}{P}{1}$\to$\term{3}{P}{2} & $E_1 \to E_0$ & 63
    & 1.70 & 0.221 & 0.022 & -0.429 & 1.63 \\
    &\term{3}{P}{0}$\to$\term{3}{P}{2} & $E_2 \to E_0$ & 44
    & 0.00446 & 2.596 & -0.235 & 1.54 & 0.0757 \\
    \hline
    ortho-\molH{} 
    &\term{3}{P}{0}$\to$\term{3}{P}{1} & $E_2 \to E_1$ & 145
    & 0.0472 & 0.393 & 0.038 & 0.230 & 29.3 \\
    &\term{3}{P}{1}$\to$\term{3}{P}{2} & $E_1 \to E_0$ & 63
    & 1.84 & 0.215 & 0.022 & 1.54 & 8.13 \\
    &\term{3}{P}{0}$\to$\term{3}{P}{2} & $E_2 \to E_0$ & 44
    & 1.23 & 0.355 & 0.012 & 1.88 & 19.1 \\
    \hline
    \end{tabular}
    \tablefoot{The form of the fits is given by Eq.~\ref{eq:Lique} and the data were taken from from \citet{Lique_2018}.}
\end{table*}

\section{Timing of \textsc{prizmo}}
\label{appendix:timing}
\begin{figure*}[ht]
    \centering
    \begin{subfigure}{0.45\linewidth}
        \includegraphics[width=\linewidth]{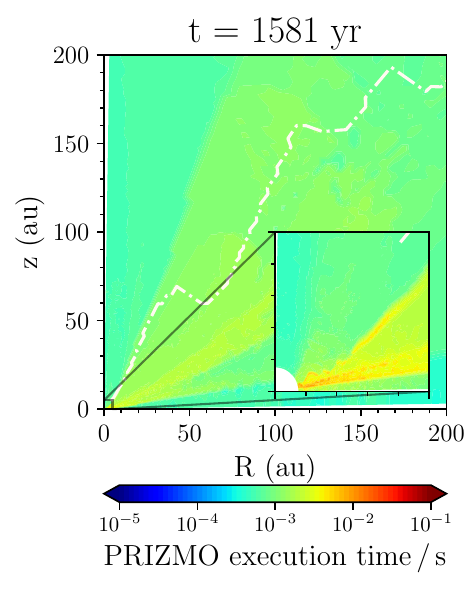}
    \end{subfigure}
    \begin{subfigure}{0.45\linewidth}
        \includegraphics[width=\linewidth]{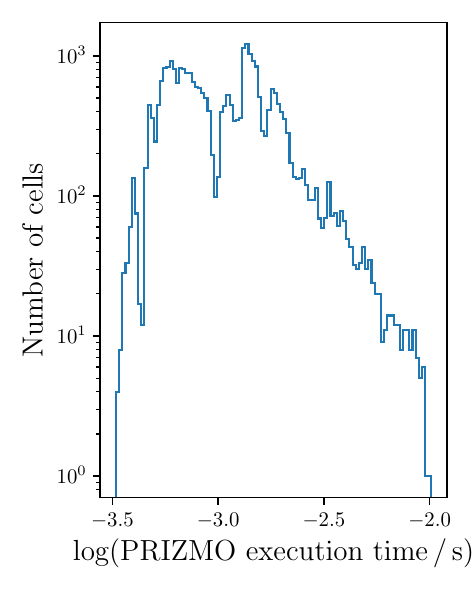}
    \end{subfigure}
    \caption{The distribution of the execution time per cell of \textsc{prizmo} after 50 orbits at 10 au.}
    \label{fig:timing}
\end{figure*}

In Fig.~\ref{fig:timing} we show information regarding the time taken by \textsc{prizmo} to calculate the thermochemistry in each cell in the simulation after 50 orbits at 10 au (the last snapshot before we freeze the simulation inside 17 au).

Generally, the most computationally expensive cells are found at heights corresponding to the disc - rather than the wind. However, the cells near the midplane are not the slowest since here we fix the temperature to the dust temperature from the DIAD models and hence the chemistry and temperature determination are decoupled.
Moreover, the longest times are seen for cells inside 10 au (as shown in the inset plot), illustrating the benefit of freezing the simulation at smaller radii.

Nevertheless, the histogram of the times shows that the distribution of the time per cell is fairly smooth and the overall time is not dominated by expensive outliers. 
Indeed since we plot $\frac{{\rm d}n}{{\rm d}\log t}$, the y-axis represents the contribution of each timescale to the overall execution time. This suggests that the large number of cells with a \textsc{prizmo} execution time $t_{\mathrm{\textsc{prizmo}}} \sim 10^{-3}\,\mathrm{s}$ dominate the total time rather than the relatively smaller number of more expensive cells. Thus, there is no easy way to decrease the total computational timescale without reducing the number of cells considerably.

Of course, since the computation is parallelised, this does not ensure that the processors are load balanced. \textsc{pluto} allows the user to choose the arrangement of the processors but not to vary the number of cells between each. Therefore, if a single processor were to contain all of the most expensive cells, these would limit the computational timescale as the other processors may have to wait a long time for this single processor to finish its calculation. To minimise the risk of this, since more expensive cells are typically found in lower latitudes or along the disc-wind interface, we try to minimise the number of cells at a particular latitude by arranging the processors to have many in the r direction since that their radial extent is large and all spend some time working on the underlying disc. While refinements may be possible, we find this gives satisfactorily little variation between the total execution time for the chemistry per processor.

\end{appendix}

\end{document}